\documentclass[nofootinbib,amsmath,amssymb,aps,twocolumn,prd,preprintnumbers]{revtex4-2}

\usepackage[colorlinks=true,linktoc=all]{hyperref}

\usepackage{graphicx}
\usepackage{dcolumn}
\usepackage{bm}

\usepackage{float}
\usepackage{xcolor}
\usepackage{bbold}
\usepackage{comment}
\usepackage{mathrsfs}

\providecommand{\abs}[1]{\lvert#1\rvert} 

\newcommand{\lagr}{\mathcal{L}}

\usepackage{cellspace} %
\newcommand{\lsim}   {\mathrel{\mathop{\kern 0pt \rlap
  {\raise.2ex\hbox{$<$}}}
  \lower.9ex\hbox{\kern-.190em $\sim$}}}
\newcommand{\gsim}   {\mathrel{\mathop{\kern 0pt \rlap
  {\raise.2ex\hbox{$>$}}}
  \lower.9ex\hbox{\kern-.190em $\sim$}}}

\begin{document}

\preprint{IPARCOS-UCM-23-130}

\title{Unified transverse diffeomorphism invariant field theory for the dark sector}

\author{David Alonso-López}
\email{dalons07@ucm.es}
\affiliation{Departamento de Física de la Tierra y
Astrofísica, Universidad Complutense de Madrid, 28040 
Madrid, Spain}

\author{Javier de Cruz Pérez}%
 \email{jadecruz@ucm.es}
\affiliation{Departamento de F\'{\i}sica Te\'orica and\\ Instituto de F\'{\i}sica de Part\'{\i}culas y del Cosmos (IPARCOS-UCM), Universidad Complutense de Madrid, 28040 
Madrid, Spain}

\author{Antonio L. Maroto}%
 \email{maroto@ucm.es}
\affiliation{Departamento de F\'{\i}sica Te\'orica and\\ Instituto de F\'{\i}sica de Part\'{\i}culas y del Cosmos (IPARCOS-UCM), Universidad Complutense de Madrid, 28040 
Madrid, Spain}

\begin{abstract}
In this work we present a unified model for the cosmological dark sector. The theory is based on a simple minimally coupled scalar field whose action only contains a canonical kinetic term and is invariant under transverse diffeomorphisms (TDiff). The model has the same number of free parameters as $\Lambda$CDM. We confront the predictions of the model at the background level with data from Planck 2018 CMB distance priors, Pantheon+ and SH0ES SNIa distance moduli, BAO data points from 6dFGS, BOSS, eBOSS and DES and  measurements of the Hubble parameter from cosmic chronometers. The model provides excellent results in the joint fit analysis, showing strong and very strong evidence compared to $\Lambda$CDM in the Bayesian evidence and Deviance Information Criterion (DIC) respectively. We also show that the Hubble tension between Planck 2018 and SH0ES measurements can be alleviated in the unified TDiff model although further analysis is still needed.
\end{abstract}

\maketitle

\section{\label{sect:intro}{Introduction}}

General Relativity's (GR) symmetry group is the group of diffeomorphisms (Diff). This symmetry arises as a consequence of Einstein's Equivalence Principle (EEP) which states geodesic motion, Lorentz invariance and Local Position Invariance (LPI) \cite{will,weinberg}; which in turn translates into the Principle of General Covariance. This principle essentially means that a physical equation is generally covariant i.e. it remains in the same form under a general coordinate transformation \cite{weinberg}. When performing an infinitesimal diffeomorphism generated by the vector field $\xi^{\mu}(x)$
\begin{equation}
    \hat{x}^{\mu}=x^{\mu}+\xi^{\mu}(x),
    \label{eqn:GCT}
\end{equation}
the metric tensor transforms with the Lie derivative i.e. $\delta g_{\mu\nu}=\mathscr{L}_{\xi}\,g_{\mu\nu}=-2\nabla_{\left(\mu\right.}\xi_{\left.\nu\right)}$. Consider now the action 
\begin{equation}
    S= S_{\text{EH}}[g_{\mu\nu}] + S_m[g_{\mu\nu},\psi^i],
\end{equation}
where 
\begin{equation}
    S_{\text{EH}}=-\frac{1}{16\pi G}\int d^4x\,\sqrt{g}\, R,
    \label{eqn:S_EH}
\end{equation}
is the Einstein-Hilbert action, $g\equiv \abs{\text{det}(g_{\mu\nu})}$ and $S_m[g_{\mu\nu},\psi^i]$ is the matter action that depends on the metric tensor as well as on the matter fields $\psi^i$. Diffeomorphism invariance implies $\delta_{\xi}S=0$, which, in turn,  implies the conservation of the energy-momentum tensor over solutions of the equations of motion of the matter fields \cite{shakespeare}, 
\begin{equation}
    \nabla_{\mu}T^{\mu\nu}=0.
    \label{eqn:consTmunu}
\end{equation}

The purpose of this work is to study the  breaking of this symmetry down to transverse diffeomorphisms (TDiff) in the matter sector. In 1919 Einstein himself was the first to introduce unimodular gravity \cite{1919}, where the determinant of the metric tensor is subject to the condition $g=1$.  As a result, the dynamical equations are the traceless Einstein's equations and the symmetry group is TDiff, i.e. diffeomorphisms subject to the condition 
\begin{equation}
    \partial_{\mu}\xi^{\mu}=0.
    \label{eqn:TDiff}
\end{equation}
The main appeal of this theory is that a cosmological constant-type term does not gravitate which could provide a solution to the the vacuum-energy problem \cite{vacuum}. Note that in these theories the symmetry is broken in the geometrical sector, however more recently the effects of Diff breaking in the matter sector have been explored in \cite{transvsobs,Maroto:2023toq,Jaramillo-Garrido:2023cor} for simple scalar field theories. There, it is shown that these models behave as ordinary
Diff  invariant theories in the geometric optics approximation, i.e. for modes
well inside the Hubble radius, where we recover the standard propagation properties of
free particles along geodesics, together with the standard scaling of energy density
for relativistic and non-relativistic particles. However for long-wavelength modes, with 
frequencies smaller than the Hubble parameter, the behaviour can be drastically different from that of the Diff models, thus opening a new avenue for  cosmological model building. 

Particularly simple cases are the minimally coupled TDiff scalar theories with 
purely kinetic terms. These models are modifications of the Diff scalar theories parametrized by a single function of the metric determinant $f(g)$. They have been seen to be equivalent to perfect adiabatic  fluids with an effective equation of state given by
 \cite{Maroto:2023toq,Jaramillo-Garrido:2023cor}
  \begin{align}
 \omega_\phi=\frac{F}{1-F},
 \end{align}
 with 
 \begin{align}
F=\frac{d\ln f}{d\ln g}, \label{F}
 \end{align}
In particular, for simple power law models $f(g)=g^\alpha$, the equation of state is just a 
constant given by 
 \begin{align}
 \omega_\phi=\frac{\alpha}{1-\alpha},
 \end{align}
which as expected, recovers the standard stiff fluid behaviour $\omega_{\phi}=1$ in the Diff case with $\alpha=1/2$. However, other choices are possible. Thus models with $\alpha=0$ would behave as non-relativistic matter fluid, not only at the background level but also 
 for perturbations, since in this simple case $c_s^2=\omega_\phi=0$ (being $c_s$ the speed of sound) providing an extremely simple model for dark matter based on a purely kinetic scalar field. 

The simplicity of these results suggests the possibility of using these kinetic TDiff models for a unified description of the dark sector with a single field. With that purpose, appropriate $f(g)$ functions which allow an interpolation between a dark matter behaviour at early times and a dark energy behaviour at late times should be identified.

The first unification models considered in the literature were based on the so called generalized Chaplygin gas \cite{Kamenshchik:2001cp}. These models are based on a single perfect fluid description of the dark sector with an equation of state given by 
\begin{align}
p=-A  \frac{1}{\rho^\alpha},
\end{align}
with $A$ a positive constant so that for $\alpha>0$ the behaviour is the expected for the 
dark sector. As a matter of fact, at the background level, the cosmological evolution
in this model allows to reduce the $H_0$ tension to the 1$\sigma$ level \cite{DiValentino:2021izs}.  The model contains one extra parameter compared to $\Lambda$CDM. When comparing the matter power spectrum with LSS data a stringent limit on the extra parameter is introduced, so that $\vert \alpha\vert \lsim 10^{-5}$, 
\cite {Sandvik:2002jz}. This constraint implies that the model effectively  behaves as $\Lambda$CDM and the solution of the tension is no longer achieved. Imperfect fluid models with bulk viscosity have also been considered for an unified description in \cite{Ren:2005nw}.

Unified models built out of a single scalar field have also been proposed in recent years.  
Thus, certain k-essence models with a Lagrangian density
\begin{align}
{\cal L}=G(X),
\end{align}
with $X=\frac{1}{2}g^{\mu\nu}\partial_\mu \phi\partial_\nu\phi$ and $G(X)$ a quadratic function of $X$ have been proposed in \cite{Scherrer:2004au}. The model avoids the problem with the matter power spectrum measurements, although the price to pay in this case is the inclusion of four extra parameters as compared to $\Lambda$CDM.

In this work we will start the analysis of  the possibilities of purely kinetic TDiff theories for the construction of viable unification models.  We will limit ourselves to the background evolution
and the comparison with  observables related to the expansion history, in particular Hubble diagrams from SNIa Pantheon+\&SH0ES, cosmic chronometers, BAO measurements and 
CMB distance priors.

This paper is organized as follows: 
in Section \ref{sect:TDiff_action} we present some general aspects of TDiff invariant actions for scalar fields, we derive the equations of motion for the field and the consistency condition that ensures conservation of the energy-momentum tensor for TDiff theories. In Section \ref{sect:Unified_model} we build the unified model for the dark sector, where we find Einstein's equations for our universe, which take the simple usual form of the Friedmann and acceleration equations. We give the expression for the energy density of the scalar field, treated as a perfect fluid and present the Hubble rates of the theory. In Section \ref{sect:Data} we present the cosmological data we used to build the likelihood to find the constraints on the parameters of the model. Section \ref{sect:Results} is devoted to present and discuss the results of the fitting process and in Section \ref{sect:Conclusions} we present our conclusions. Finally, in the Appendix we show the basic derivation of the TDiff invariant action for a scalar field.

Throughout this manuscript we will use the metric signature $(+,-,-,-)$ and natural units $\hbar=c=1$. \footnote{The Hubble parameter $H_0$ will be expressed sometimes in units of km/s/Mpc to ease the comparison with other results in the literature where this convention is commonly used.}

\section{TDiff invariant action for a scalar field}\label{sect:TDiff_action}

Let us start by writing the TDiff invariant action for a simple minimally coupled scalar field in the kinetic regime (see Appendix for details)
\begin{align}\label{eq:action_with_kinetic_term}
  S&= S_{\text{EH}}+ S_{\phi}\\
  &=-\frac{1}{16\pi G}\int d^4x \sqrt{g}\,R + \int d^4x\, f(g)\, \frac{1}{2}g^{\mu\nu}\partial_{\mu}\phi\partial_{\nu}\phi,\nonumber
\end{align}
where $f(g)$ is a positive function of the metric determinant.  This guarantees a stable theory which respects the Weak Equivalence Principle in the
geometric optics approximation \cite{Maroto:2023toq}. The gravitational sector is described by the standard Einstein-Hilbert action
so that only the scalar sector breaks Diff invariance down to TDiff\footnote{See \cite{Bello-Morales:2023btf} for a detailed study of gravity models which break Diff invariance in the geometrical part of the action.}. Notice that the standard Diff invariant theory is recovered in the case $f(g)=\sqrt{g}$.

\subsection{TDiff models in cosmological backgrounds}

We now would like to specify the geometry in order to build cosmological models for homogeneous scalar fields, i.e. coupled to a homogeneous and isotropic spacetime such as Robertson-Walker (RW). In this cosmological case, in which we have a spacetime whose constant time hypersurfaces are maximally symmetric subspaces, symmetry implies that, in the general case, only 2 metric components are truly independent, up to the sign of the curvature of these hypersurfaces. We will work with flat spatial sections, so the line element of the most general homogeneous and isotropic spacetime is \cite{weinberg}
\begin{equation}
    ds^2=b^2(\tau)d\tau^2-a^2(\tau)d\vec{x}^2,
    \label{eqn:RW}
\end{equation}
where $a(\tau)$ and $b(\tau)$ are the two independent components of the metric. 

It is most common to define a new time coordinate, known as cosmological time $dt=b(\tau)d\tau$ which finally leaves us with only one function $a(t)$ to solve for. Note that this is only possible in the case of a full Diff theory. In general there is no way to perform a coordinate transformation satisfying the TDiff condition that allows us to write \eqref{eqn:RW} using cosmological time $t(\tau)$. 

\subsection{Equations of motion for $\phi(\tau)$}

The scalar field equations of motion obtained from \eqref{eq:action_with_kinetic_term}
read
\begin{align}
\partial_\nu(f(g) g^{\mu \nu}\partial_\mu\phi)=0\label{KGm}.
\end{align}

Let us consider the evolution of a homogeneous scalar field only depending on time, so that the equations of motion satisfied by $\phi(\tau)$ in the geometry \eqref{eqn:RW} are \cite{Maroto:2023toq}
\begin{equation}
    \phi''+\left(\frac{f'}{f}-2\frac{b'}{b}\right)\phi'=\phi''+\frac{L'}{L}\phi'=0,
    \label{eqn:eom_tau}
\end{equation}
where the prime denotes the derivative with respect to the coordinate time $\tau$, and 
\begin{align}
    L(\tau)=\frac{f(g(\tau))}{b^2(\tau)}.
\end{align}
 It is easy to check that in the $f(g)=\sqrt{g}$ case, we recover the well-known results for a Diff invariant scalar field. 
 Equation \eqref{eqn:eom_tau} can be rewritten as a total derivative 
\begin{equation}
    \frac{d}{d\tau}(L\phi')=0,
\end{equation}
so one immediately obtains \cite{Maroto:2023toq}
\begin{equation}
    \phi'(\tau)=\frac{C_{\phi}}{L(\tau)}.
\end{equation}
with $C_\phi$ a constant.

 Even though it is not possible to set
 $b=1$ with a TDiff coordinate transformation, we can always rewrite \eqref{eqn:eom_tau} in 
 cosmological time $dt=b(\tau)d\tau$ so that we obtain
\begin{align}
\ddot \phi+\frac{\dot J}{J}\dot \phi=0\label{KGRW}
\end{align}
where
\begin{align}
J=Lb=\frac{f}{b}=a^3\frac{f}{\sqrt g}
\end{align}
and dot denotes derivative with respect to the cosmological time $t$.

\subsection{Einstein's equations and the energy-momentum tensor}

The field equations derived from the action \eqref{eq:action_with_kinetic_term} with 
unconstrained variations with respect to the metric tensor
\begin{align}
    -\frac{2}{\sqrt{g}}\frac{\delta S}{\delta g_{\mu\nu}}=0, 
\end{align}
are the Einstein's equations
\begin{equation}
    G_{\mu\nu}=8\pi G \,
    T^\phi_{\mu\nu},
    \label{eqn:einsteineqs}
\end{equation}
where
\begin{equation}
    T^\phi_{\mu\nu}=\frac{f(g)}{\sqrt{g}}\left(\partial_{\mu}\phi\partial_{\nu}\phi-F(g)g_{\mu\nu}g^{\alpha\beta}\partial_{\alpha}\phi\partial_{\beta}\phi\right).
    \label{eqn:Tmunu_phi}
\end{equation}
with $F(g)$ given in \eqref{F}. 

Notice that because of the breaking of Diff invariance in the scalar sector, $T^\phi_{\mu\nu}$
is not necessarily conserved on solutions of the scalar field equations of motion. 
In spite of that, since Einstein's equations \eqref{eqn:einsteineqs} still hold, Bianchi identities imply the conservation of the energy-momentum tensor \eqref{eqn:Tmunu_phi} over solutions of the Einstein's equations.

Since our goal is to build a cosmological model, we would like to interpret our field as a perfect fluid\footnote{See \cite{Jaramillo-Garrido:2023cor} for a discussion of the condition to be met in order to describe the scalar field component that breaks Diff invariance down to TDiff as a perfect fluid.}, whose energy-momentum tensor is well-known to be parametrized as
\begin{equation}
    T_\phi^{\mu\nu}=\rho_\phi\, u^{\mu}u^{\nu}-p_\phi\, h^{\mu\nu},
    \label{eqn:ideal}
\end{equation}
where $u^{\mu}$ is the 4-velocity of the fluid, and as such it must obey the normalization condition $g_{\mu\nu}u^{\mu}u^{\nu}=1$, and $h_{\mu\nu}=g_{\mu\nu}-u_{\mu}u_{\nu}$ is the metric tensor on the hypersurfaces orthogonal to the direction of the time-like vector $u^{\mu}$. We can read the energy density $\rho_\phi$ and the pressure $p_\phi$ of our fluid to be
\begin{eqnarray}
    \rho_\phi=T_\phi^{00}g_{00}&=&\frac{f(g)}{b^2\sqrt{g}}\left(1-F(g)\right)\phi'^2,
    \label{eqn:rho} \\
    p_\phi=-\frac{T_{\phi{i}}^i}{3}&=&\frac{f(g)}{b^2\sqrt{g}}F(g)\phi'^2.
    \label{eqn:presion}
\end{eqnarray}
From these two expressions we are able to define an equation of state for the fluid in the usual way $p_\phi(a)=\omega_{\phi}\rho_\phi(a)$, where
\begin{equation}
    \omega_{\phi}=\frac{F}{1-F}. 
    \label{eqn:w}
\end{equation}
Notice that the equation of state parameter solely depends on our choice of $f(g)$. It is worth mentioning that the problematic case $F=1$ is excluded since it implies $\rho_\phi=0$ in \eqref{eqn:rho}. It is already possible to draw some conclusions on how the fluid behaves from \eqref{eqn:w}. If we had a power law $f\propto g^{\alpha}$, then 
\begin{equation}
    \omega_{\phi}=\frac{\alpha}{1-\alpha},
\end{equation}
which is time independent. In order to have positive energy density we must have $\alpha < 1$. To recover a non-relativistic/dark matter type behaviour we simply need $\alpha=0$, whereas $\alpha\rightarrow\infty$ implies a cosmological constant behaviour $\omega_{\phi}=-1$. We can interpolate between these two limiting cases with an exponential function $f(g)\propto e^{-\beta g}$, resulting in
\begin{equation}
    \omega_{\phi}=-\frac{\beta g}{1+\beta g},
    \label{eqn:wphi}
\end{equation}
where $\beta g\ll 1 \Rightarrow \omega_{\phi}=0$ whilst $\beta g\gg 1 \Rightarrow \omega_{\phi}=-1$. The condition required for a positive energy density is now $\beta g>-1$ \cite{Maroto:2023toq}.

Computing $\nabla_{\mu}T^{\mu\nu}_\phi$ yields\footnote{We have used the fact that for a homogeneous and isotropic geometry there are no pressure gradients at a given space-like hypersurface of $\tau=\text{const}$, i.e. $h^{\mu\nu}\nabla_{\mu}p=0$; which ensures geodesic motion $u^{\mu}\nabla_{\mu}u^{\nu}=0$.}
\begin{equation}
\begin{split}
    \nabla_{\mu}T_\phi^{\mu\nu}&=u^{\nu}(u^{\mu}\nabla_{\mu}\rho_\phi+(\rho_\phi+p_\phi)\nabla_{\mu}u^{\mu})=\\[5pt]
        &=\frac{u^{\nu}}{b}\left(\rho^{\prime}_\phi+3\,\frac{a'}{a}(\rho_\phi+p_\phi)\right)=0.
\end{split}
\end{equation}
As expected for a RW geometry, there is actually just one independent equation\footnote{Note that despite the fact that $b'$ does not explicitly appear on \eqref{eqn:continuity}, the energy density and the pressure of the fluid do depend on $b(\tau)$, and hence a dependence on $b'$ is implicit in $\rho'_{\phi}$.} 
\begin{equation}
    \rho^{\prime}_\phi+3\,\frac{a'}{a}(\rho_\phi+p_\phi)=0. 
    \label{eqn:continuity}
\end{equation}
The compatibility of Eq. \eqref{eqn:continuity} and the equations of motion of the field \eqref{eqn:eom_tau} will impose a condition on $b(\tau)$. Though not always possible to find such a condition in an explicit way, in the kinetic domination regime it can actually be realized \cite{Maroto:2023toq}. By expressing $\rho_\phi'$ in terms of $\rho_\phi$, which requires the use of the equations of motion \eqref{eqn:eom_tau}, since $\rho_\phi+p_\phi=\rho_\phi/(1-F)$ we get rid of the energy density explicit dependence and only terms involving the geometry $b$, $a$, and $f(g)$ will appear. After some manipulations we find
\begin{equation}
    \frac{g'}{g}\left(F-\frac{1}{2}\right)+\frac{F'}{1-F}=6\,\left(\frac{a'}{a}\right)\,\frac{F-\frac{1}{2}}{1-F}, 
\end{equation}
which is written in such a way that is trivially satisfied when $F=1/2 \Rightarrow F'=0$. Multiplying both sides by $(1-F)/(F-1/2)$ we finally arrive at\footnote{Note that this expression no longer recovers the GR identity as a consequence of implicitly dividing by zero when $F=1/2$ in the last step of the derivation.}
\begin{equation}
    \frac{g'}{g}(1-F)-\frac{2F'}{1-2F}=6\,\frac{a'}{a}.
    \label{eqn:pre_clave}
\end{equation}
Integration yields \cite{Maroto:2023toq}
\begin{equation}
    \frac{g}{f(g)}(1-2F(g))=C_g\,a^6, 
    \label{eqn:CLAVE}
\end{equation}
with $C_g$ a constant.
For TDiff invariant models, we must solve this equation for a given $f(g)$ in order to have a conserved energy-momentum tensor. This will give us a relation between the two functions $b(\tau)$ and $a(\tau)$ i.e. $b(a)$.

\section{A unified model for the dark sector}\label{sect:Unified_model}

Once we have discussed what it means to have a TDiff invariant action for a scalar field, and what conditions must be satisfied in order to build a consistent model with it, we shall put this knowledge to use to construct a TDiff invariant theory for the dark sector in cosmology. We would like to extend the action in \eqref{eq:action_with_kinetic_term} to also account for the well-known physics, that is, the particles of the Standard Model. To do so, we proceed as usual by including a full Diff action for baryonic matter (B) and photons plus massless neutrinos, which will be referred to as radiation (R), all of them treated as perfect fluids.

So, how to model dark matter and dark energy with a single TDiff invariant scalar field? As discussed in last section, we saw that for an exponential function 
\begin{align}
f(g)= e^{-\beta g} \label{f}
\end{align}
we could interpolate between a matter-type behaviour at early times ($\omega_{\phi}(a\to 0)=0$) and a cosmological constant-type one ($\omega_{\phi}(a\to 1)=-1$) at late times. This will be our choice since it allows for a unified description for both dark matter and dark energy. The action reads
\begin{equation}\label{eqn:Total_action}
S=S_{\text{EH}}+S_{\textrm{B+R}}+S_{\phi},
\end{equation}
where $S_{\text{EH}}$ is given in \eqref{eqn:S_EH}, $S_{\textrm{B+R}}$ takes into account the baryon and radiation contributions, and
\begin{equation}
    S_{\phi}=\int d^4x\, e^{-\beta g}\, \left(\frac{1}{2}g^{\mu\nu}\partial_{\mu}\phi\partial_{\nu}\phi\right),
    \label{eqn:Sphi_model}
\end{equation}
for the TDiff scalar sector. 
This means that Einstein's equations are
\begin{equation}
    G_{\mu\nu}=8\pi G\, (T^{\textrm{B+R}}_{\mu\nu}+T^{\phi}_{\mu\nu}), 
\end{equation}
where $T^{\textrm{B+R}}_{\mu\nu}$ corresponds to the energy-momentum tensor of baryons and radiation and $T^{\phi}_{\mu\nu}$ represents the contribution from the dark sector. As a good approximation we can consider that baryons and radiation do not interact at the background level with each other which means that both are self-conserved. In the case of baryons the energy density evolves according to $\rho_B(a) = \rho_B{a^{-3}}$ \footnote{When we do not indicate the explicit dependence in the scale factor or in the redshift, for the energy densities we are referring to the corresponding present value.} whereas its equation of state is $\omega_B = 0$. The energy density, as a function of the scale factor, of radiation can be written in terms of photon's energy density ($\rho_\gamma(a)$) as follows
\begin{equation}
\rho_R(a) = \left(1 + \frac{7}{8}N_{\textrm{eff}}\left(\frac{4}{11}\right)^{4/3}\right)\rho_\gamma(a),    \label{radiation} 
\end{equation}
where $N_{\textrm{eff}}=3.046$ is the effective number of neutrino species and $\rho_\gamma(a) = \rho_\gamma{a^{-4}}$. The equation of state for the radiation fluid is $\omega_R = 1/3$.
As for the dark sector we can take advantage of the novel result obtained in \cite{Jaramillo-Garrido:2023cor}. In this reference it is shown that, as long as the derivative of the scalar field $\partial_\mu\phi$ is a time-like vector, from the conservation of the energy-momentum tensor, in the kinetic domain, the energy density associated to the scalar field can be expressed as follows 
\begin{equation}\label{eq:rho_SF_2}
\rho_\phi(a) = \frac{C_\phi}{(\omega_\phi -1)\sqrt{g}}, 
\end{equation}
where $C_\phi$ is a constant that needs to be fixed so that the cosmic sum rule is fulfilled.

The two independent components of the Einstein tensor are
\begin{eqnarray}
    G_{00}&=&3\left(\frac{a'}{a}\right)^2, \\
    G_{ii}&=&-2\left(\frac{a}{b}\right)^2\left(\frac{a''}{a}-\frac{a'}{a}\frac{b'}{b}+\frac{1}{2}\left(\frac{a'}{a}\right)^2\right).
\end{eqnarray}
Using $T_{00}=b^2\left[\rho_B(a) + \rho_R(a) +\rho_\phi(a)\right]$ and $T_{ii}=a^2\left[p_R(a) +p_{\phi}(a)\right]$ we get
\begin{align}
    &\left(\frac{a'}{a}\right)^2=\frac{8\pi G}{3}\left[\rho_B(a)+ \rho_R(a) +\rho_{\phi}(a)\right]\,b^2, \\
    &\frac{a''}{a}-\frac{a'}{a}\frac{b'}{b}+\frac{1}{2}\left(\frac{a'}{a}\right)^2=-4\pi G\, \left[p_R(a)+p_{\phi}(a)\right]\, b^2.
\end{align}

As we have mentioned before, if we had set $b=1$ from the beginning,  no simultaneous solutions for the field equations \eqref{eqn:eom_tau} and the conservation of energy \eqref{eqn:CLAVE} could have been found \cite{Maroto:2023toq}. However, once we have obtained the equations of motion, it is possible to perform a change of variables to  cosmological time $dt=b(\tau)d\tau$. After doing so we obtain the equivalent to the usual Friedmann and acceleration equations that take the standard form
\begin{align}
    &H^2=\frac{8\pi G}{3}\left[\rho_B(a) + \rho_R(a)+\rho_{\phi}(a)\right],
    \label{eqn:Friedmann}\\
    &\frac{\ddot a}{a}=-\frac{4\pi G}{3}\left[\rho_B(a) + \rho_R(a)+\rho_{\phi}(a) + 3p_R(a)+3p_{\phi}(a)\right]
    \label{eqn:aceleracion},
\end{align}
where we have used \eqref{eqn:Friedmann} to obtain \eqref{eqn:aceleracion} and $H=\dot a/a$ is just the Hubble rate. Notice that although at first glance it might look as if there was no explicit $b(t)$ dependence, the energy density and pressure of the dark sector do depend on it. It is there were the consistency relation $b(a)$ \eqref{eqn:CLAVE} comes into play. Imposing $f(g)=e^{-\beta g}$ and $F(g)=-\beta g$ it can be written as
\begin{equation}\label{eq:constraint_2}
b^2{e^{\beta{a^6}b^2}}\left(1 +2\beta{a^6}b^2\right) = C_g.    
\end{equation}
We can fix the value of the constant $C_g$ by evaluating the left hand side at present time, namely $a = b =1$, which automatically implies that the determinant of the metric is normalized as $g(a=1)=1$. We end up with the following expression for the constraint
\begin{equation}\label{eqn:b(a)}
{b}^2{e^{{\beta}(a^6{{b}^2} -1)}}\frac{\left(1+2{\beta}a^6{{b}^2}\right)}{1 + 2{\beta}} = 1.     
\end{equation}
Thanks to the above expression we will be able to get $b = b(a,\beta)$.  

The Hubble rate can be obtained from \eqref{eqn:Friedmann}
\begin{equation}
    H(a)=H_0\sqrt{\Omega_B\,a^{-3}+ \Omega_R\,a^{-4}+ \Omega_{\phi}(a)},
    \label{eqn:Hubble}
\end{equation}
where we have considered the definitions $\Omega_i(a) \equiv \frac{\rho_i(a)}{\rho_c}$ \footnote{It is important to note that with this definition the cosmic sum rule is only fulfilled at present time.}, with $i=\text{B,R},\phi$ and being $\rho_c=3H_0^2/(8\pi G)$ the value today of the critical energy density. 

The cosmic sum rule evaluated at present time reads
\begin{equation}\label{eq:sum_rule}
    1=\Omega_B+ \Omega_R + \Omega_{\phi}.
\end{equation}
As stated before $C_\phi$ in \eqref{eq:rho_SF_2} can be fixed with relation \eqref{eq:sum_rule}. By doing so we get
\begin{equation}\label{eqn:phi_energy_density}
\Omega_\phi(a) = \Omega_\phi\left(\frac{\omega_\phi(a=1) -1}{\omega_\phi(a) -1}\right)\frac{1}{\sqrt{g}}.    \end{equation}

Once we have \eqref{eqn:b(a)} and \eqref{eqn:phi_energy_density} we can study the high redshift limit. From \eqref{eqn:b(a)} we get the following approximation $b \simeq \sqrt{1+2\beta}e^{\frac{\beta}{2}}$ which in turn allows us to find an approximate expression for the energy density of the scalar field valid at short times
\begin{equation}\label{eq:Omega_phi_a_small}
\Omega_\phi(a) \simeq  \Omega^{\textrm{eff}}_{\textrm{DM}}{a^{-3}} ,    
\end{equation}
where we have used
\begin{equation}\label{eq:Omega_phi_eff}
\Omega^{\textrm{eff}}_{\textrm{DM}} \equiv \Omega_{\phi}e^{-\frac{{\beta}}{2}}\frac{\sqrt{1+2{\beta}}}{1+{\beta}}.    
\end{equation}
This does not come as a surprise since as we have stated previously when $\beta{g}\ll 1$ the equation of state tends to the value $\omega_\phi = 0$, therefore it behaves as non-relativistic matter.

\begin{figure}[t]
    \centering
    \includegraphics[scale=0.6]{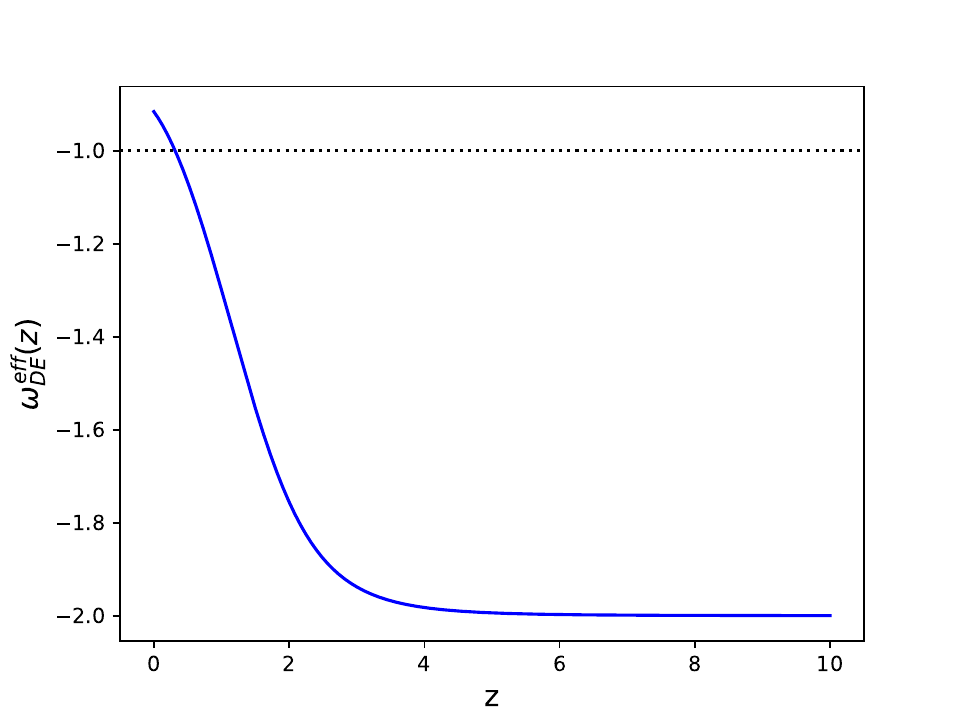}
    \caption{The effective equation of state parameter of the dark energy component for the TDiff model \eqref{eq:w_DE_TDiff}. The values of the cosmological parameters can be seen in Table \ref{tab:main_results}. }
    \label{fig:w_DE_TDiff}
\end{figure}

\begin{figure}[t]
    \centering
    \includegraphics[scale=0.6]{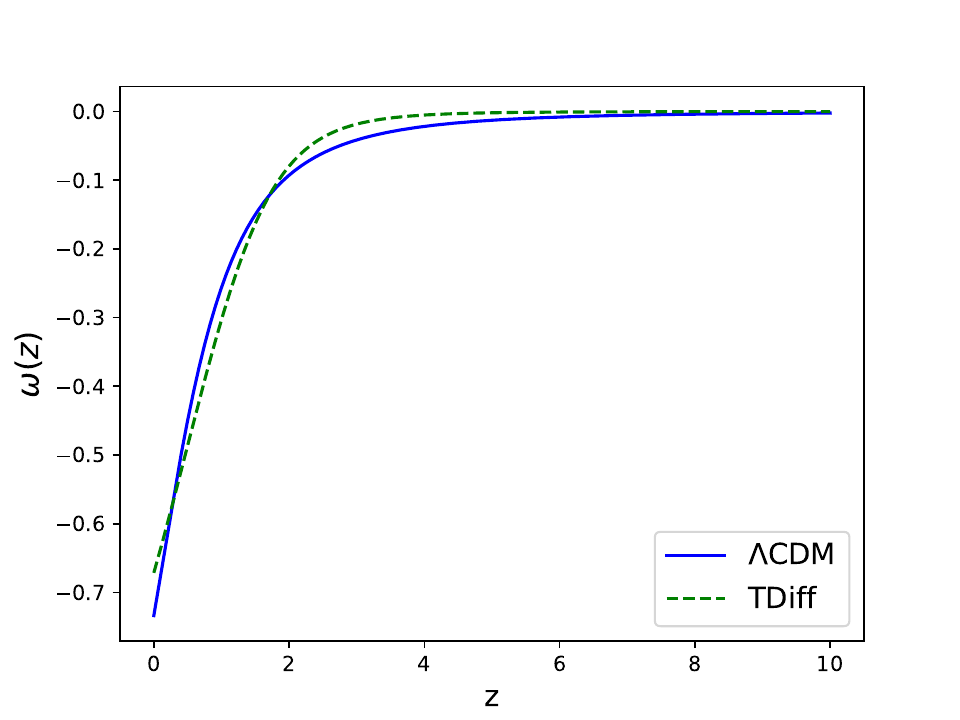}
    \caption{Comparison of the equation of state for the dark sector. For the $\Lambda$CDM $\omega_{\Lambda\text{CDM}}(z) = (p_{\textrm{DM}}(z) + p_\Lambda)/(\rho_{\textrm{DM}}(z) + \rho_\Lambda)$ whereas for the TDiff model $\omega_\phi(z)$ is written in \eqref{eqn:wphi}. The values of the cosmological parameters can be seen in Table \ref{tab:main_results}.}
    \label{fig:eqn_state}
\end{figure}

At the background level we can separate the contribution of the scalar field as if it were composed of an effective dark matter contribution, whose expression is determined by Eq. \eqref{eq:Omega_phi_a_small} and an effective dark energy component:   
\begin{align}
&\rho_\phi(a) = \rho^{\textrm{eff}}_{\textrm{DM}}(a) + \rho^{\textrm{eff}}_{\textrm{DE}}(a) = \rho_{c}\Omega^{\textrm{eff}}_{\textrm{DM}}{a^{-3}} +  \rho^{\textrm{eff}}_{\textrm{DE}}(a), \\ 
& p_\phi(a) = p^{\textrm{eff}}_{\textrm{DM}}(a) + p^{\textrm{eff}}_{\textrm{DE}}(a) = \omega^{\text{eff}}_{\textrm{DE}}(a)\rho^{\textrm{eff}}_{\textrm{DE}}(a).
\end{align}
As stated previously, due to the form of the equation of state parameter $\eqref{eqn:wphi}$ the scalar field behaves as non-relativistic matter at early times and as dark energy at late times. Therefore, considering the above equations we can tell when the effective dark energy component has a quintessence-like behaviour and when it has a phantom-like behaviour. The expression for the equation of state parameter is the following one
\begin{equation}\label{eq:w_DE_TDiff}
\omega^{\textrm{eff}}_{\textrm{DE}}(a) = \omega_\phi(a)\frac{\rho_\phi(a)}{\rho_\phi(a) -\rho^{\textrm{eff}}_{\textrm{DM}}(a)}.    
\end{equation}
As an example, in Fig. \ref{fig:w_DE_TDiff} we plot its evolution for the values of the cosmological parameters in Table \ref{tab:main_results}. As it can be seen, $\omega^{\text{eff}}_{\text{DE}}(a)\simeq-2$, at high redshift, something that can be checked from the analytical expression \eqref{eq:w_DE_TDiff}. This asymptotic value is determined by the particular form of the $f(g)$ function in \eqref{f} regardless the specific value
of $\beta$. The value of $\omega^{\text{eff}}_{\text{DE}}(a)$ remains in the phantom region for most of the cosmic history, however, as we get closer to present time its value starts to increase until it crosses the phantom divide. This is the opposite of other unified models such as  the generalized Chaplygin gas  in which no crossing of the phantom divide line occurs in the effective dark energy equation of state. The crossing of the phantom divide line has been shown in \cite{Heisenberg:2022gqk} to be  a necessary condition to solve the $H_0$ and $\sigma_8$ tensions in models of dark energy.

Despite the different behaviour of the TDiff effective dark energy
equation of state with respect to a cosmological constant, the full dark sector behaves similarly to $\Lambda$CDM. 
Thus in Fig. \ref{fig:eqn_state} we compare the  dark sector equation of state given by   
\begin{align}
    \omega_{\Lambda\text{CDM}}(z) = \frac{p_{\textrm{DM}}(z) + p_\Lambda}{\rho_{\textrm{DM}}(z) + \rho_\Lambda} \label{wLCDM}
\end{align}
in $\Lambda$CDM
 with $\omega_\phi(z)$ written in \eqref{eqn:wphi} for the TDiff model for the cosmological parameters in
Table \ref{tab:main_results}.

\subsection{Preferred coordinate time}
\label{sec:Hubble_time_rate}

\begin{figure}[t]
    \centering
    \includegraphics[scale=0.6]{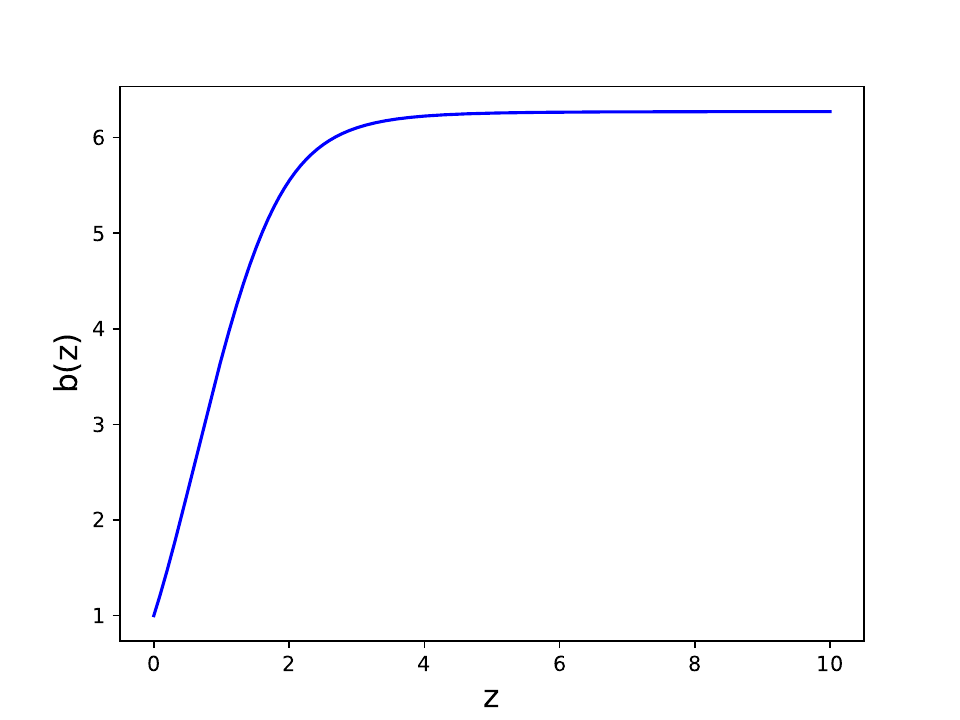}
    \caption{Time scale factor $b$ as a function of redshift $z=1/a-1$ given by the consistency relation \eqref{eqn:b(a)} that follows from the conservation of the energy-momentum tensor and the equations of motion of the field. The values of the cosmological parameters can be found in Table \ref{tab:main_results}.}
    \label{fig:b_param}
\end{figure}

\begin{figure}[t]
    \centering
    \includegraphics[scale=0.6]{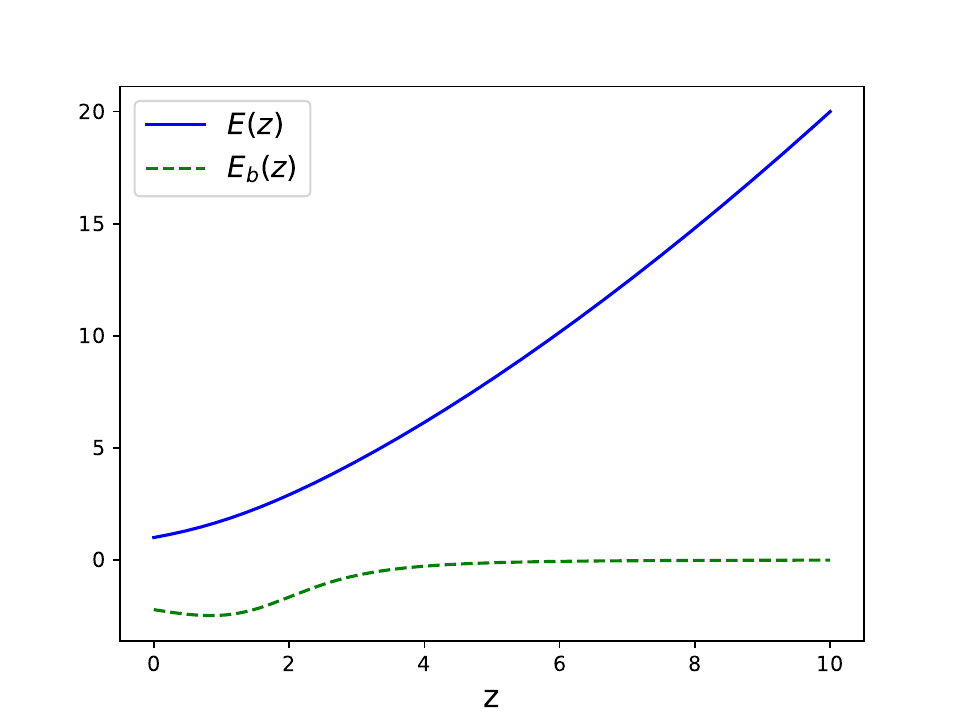}
    \caption{The normalized Hubble rate $E(z)=\frac{1}{H_0}\frac{\dot a}{a}$ and $E_b=\frac{1}{H_0}\frac{\dot b}{b}$. The values of the cosmological parameters can be found in Table \ref{tab:main_results}.  }
    \label{fig:Hubble_TDiff}
\end{figure} 
A unique feature arising from the breaking of the full Diff symmetry is the existence of a preferred coordinate time. In General Relativity the lapse function $b(\tau)$ is a pure gauge component in cosmological space-times which can be set to $b(t)=1$ with a time coordinate transformation. However, in TDiff models $b(\tau)$ is a physical metric component which sets the privileged coordinate time which 
is related to the cosmological time (that measured by comoving observers)
by $b=\frac{dt}{d\tau}$. We can interpret this time scale factor as the rate of change of cosmological time with respect to the privileged time $\tau$, so that, whenever $b$ is constant, means that the two times are essentially the same up to a constant factor. 
In Fig. \ref{fig:b_param} we show $b(z)$ for the TDiff model with 
the cosmological parameters in Table \ref{tab:main_results}. We see the  constant behaviour from the early universe up until the beginning of the accelerated expansion, where it starts to quickly decrease. 
Notice that the preferred 
time is unique up to time translations since no time reparametrization was possible via a coordinate transformation satisfying the TDiff condition \eqref{eqn:TDiff} while maintaining the spatial coordinates that ensure homogeneity and isotropy.

We can now define the analogue to the usual (space) Hubble rate $H\equiv\frac{\dot a}{a}$, but for the lapse function $b(t)$ i.e. a time Hubble rate
\begin{equation}
   H_b(a)\equiv\frac{\dot b}{b}=\frac{d\ln{b}}{d\ln{a}}\,H(a)=B(a)H(a), 
\end{equation}
where $B(a)$ can be written in terms of the following combination of $F(g)$
\begin{equation}
    B(a)=\frac{3F(3+2F)}{1-F(5+2F)}.
\end{equation}

We find that both Hubble rates are related via the function $B(a)\equiv\frac{a}{b}\, \frac{db}{da}$. This is a consequence of the consistency relation \eqref{eqn:CLAVE}, which essentially implies that the time scale factor is a function of the spatial scale factor. We then conclude that $b$ cannot evolve with time unless $a$ does. Nonetheless this is not necessarily true the other way around, as seen in Fig. \ref{fig:b_param}.

Fig. \ref{fig:Hubble_TDiff} shows both Hubble rates in units of $H_0$ as a function of redshift. $H(z)$ decreases with time whereas $H_b(z)$ goes to zero at large redshifts, so $b(z)$ is constant at the matter dominated era, which is consistent with the discussion above. This is in agreement with the fact that at early times $f(g)=e^{-\beta g}\simeq g^0$, so it behaves as a power law for which $b(a) \propto a^{\frac{3\alpha}{1-\alpha}}$ with $\alpha=0$, so $b\simeq \text{const}.$ \cite{Maroto:2023toq}. The fact that $H_b$ is negative implies that $B(a)<0$.
%
%
\section{Data and methodology}\label{sect:Data}

The model we have just introduced contains the same number of free parameters as $\Lambda\text{CDM}$. Thus, in the flat $\Lambda\text{CDM}$, we have at the background level, three free parameters: $(\Omega_M,\Omega_B,H_0)$ where $\Omega_M = \Omega_B + \Omega_{\textrm{DM}}$ contains the contribution from baryons and dark matter\footnote{We consider that the radiation contribution is precisely known from the CMB temperature measurement and the effective number of neutrino species and is given in \eqref{radiation}. Since we work in natural units the Hubble parameter is defined by $H_0^{-1}=2997.9\, h^{-1}\, \mathrm{Mpc}^{-1}$.}. As for the TDiff cosmological model, we also have three free parameters $(\beta,\Omega_B,H_0)$, where the parameter $\beta$, somehow, plays a similar role to $\Omega_M$ in $\Lambda$CDM (see \eqref{eq:Omega_phi_eff}). It is important to note that in addition to the main cosmological parameters that we have just mentioned we also have to consider the nuisance parameter $M$ which stands for the absolute magnitude in the supernovae data. We shall provide more details about it when we describe the corresponding likelihood. In the following we describe the cosmological data sets employed in this work. 

\textbf{CMB:} Instead of considering the full likelihood for Planck 2018 data here we use the CMB distance priors, which for those models that do not depart too much from the $\Lambda$CDM predictions at high redshift, turns out to be an efficient way of compressing the information (see section 5.1.6 of \cite{Planck:2015bue} for more details). The two distance priors that we utilize here are the shift parameter $R$ and the acoustic length $\ell_a$ that can be computed from 
\begin{align}
R &= \sqrt{\Omega_{m}H^2_0}(1+z_*)D_A(z_*), \\ 
\ell_a &= {\pi}(1+z_*)\frac{D_A(z_*)}{r_s}.
\end{align}
The above expressions contain the angular distance 
\begin{equation}
D_A(z) = \frac{1}{(1+z)}\int^z_0 \frac{dz^\prime}{H(z^\prime)},  
\end{equation}
evaluated at the decoupling redshift $z_*$, which depends on the free cosmological parameters. In order to get its value we consider the fitting formulae presented in \cite{Hu:1995en}: 
\begin{align}
& z_{*} = 1048\left[1+0.00124\omega^{-0.738}_b\right]\left[1+ g_1\omega^{g_2}_m\right],\\
& g_1 = \frac{0.0783\omega^{-0.238}_b}{1+39.5\omega^{0.763}_b},\\
& g_2 = \frac{0.560}{1+ 21.1\omega^{1.81}_b},
\end{align}
where $\omega_b \equiv \Omega_B{h^2}$ and $\omega_m \equiv \Omega_M{h^2}$. For the TDiff model we do not have the parameter $\Omega_M$, however, since the above formulas are meant to compute $z_{*}\sim\mathcal{O}(10^3)$, we are allowed to use \eqref{eq:Omega_phi_eff} hence $\Omega^{\text{eff}}_M \equiv \Omega^{\text{eff}}_{\text{DM}} + \Omega_B$. The term $r_s$ stands for the radius of the sound horizon 
\begin{equation}\label{eqn:sound_horizon}
r(z) = \int^{\infty}_{z}\frac{c_s(z^\prime)dz^\prime}{H(z^\prime)},
\end{equation}
evaluated at the decoupling epoch $r_s\equiv r(z_*)$ being $c_s(z)$ the speed of sound in the photon-baryon fluid. In addition to the two distance priors we also consider the value of the baryon density $\omega_b$. For the Planck 2018 TT,TE,TE+lowE+lensing \cite{Planck:2018vyg} data we obtain: 
\begin{align}\label{eq:CMB_data_points}
& \omega_b = 0.02239\pm 0.00015,\\
& \ell_a = 301.529\pm 0.083,\\
& R = 1.7497 \pm 0.0041, 
\end{align}

with the corresponding correlation matrix 
\begin{small}
\begin{equation}\label{eq:correlationCMB}
C_{\textrm{CMB}} = 
\begin{pmatrix}
1 & -0.2164 & -0.6417   \\
-0.2164 & 1 & 0.3553    \\
-0.6417 & 0.3553 & 1
\end{pmatrix}.
\end{equation}
\end{small} 

\textbf{BAO:} From isotropic and anisotropic estimators we consider 12 data points, probing the redshif range $0.122\leq z \leq 2.334$. Here we consider the likelihoods without the growth rates points. The list of measurements utilized in our analysis (for the sake of convenience in some cases we do not show all the decimals of the measurements and the covariance matrices though in the analysis all of them have been considered) can be seen in Table \ref{tab:bao}, and the quantities that appear are, the transverse comoving distance
\begin{equation}
D_M(z) = (1+z)D_A(z),    
\end{equation}
the Hubble radius 
\begin{equation}
D_H(z) = \frac{1}{H(z)},    
\end{equation}
and the angle-averaged distance
\begin{equation}
D_V(z) = \left[zD_M^2(z)/H(z)\right]^{1/3} .   
\end{equation}
As usual the observational data points obtained from the BAO analysis are provided as relative distances with respect to the radius of the sound horizon evaluated at the drag epoch $r_d\equiv r(z_d)$. In order to compute the value of $z_d$, which depends on the value of the cosmological parameters, we make use of the fitting formulae provided in \cite{Hu:1995en}:
\begin{align}
& z_d = 1345\frac{\omega^{0.251}_m}{1+0.659\omega^{0.828}_m}\left[1+ b_1\omega^{b_2}_{b}\right],\\
& b_1 = 0.313\omega^{-0.419}_m\left[1+0.607\omega^{0.674}_m\right],\\
& b_2 = 0.238\omega^{0.223}_m.
\end{align}
Finally we provide the corresponding covariance matrices.

\begin{widetext}
The covariance matrix between measurements for the BOSS Galaxy data \cite{Gil-Marin:2020bct} is given by
\begin{equation}
   \textrm{Cov}_\textrm{BOSS-Galaxy}=
   \begin{pmatrix}
      0.02860520 & -0.04939281 & 0.01489688 & -0.01387079  \\
      -0.04939281 & 0.5307187 & -0.02423513 & 0.01767087 \\
      0.01489688 & -0.02423513 & 0.04147534 & -0.04873962 \\
      -0.01387079 & 0.01767087 & -0.04873962 & 0.3268589 
   \end{pmatrix} .
\end{equation}
\end{widetext}

For the two data-points obtained with eBOSS LRG data \cite{Gil-Marin:2020bct,Bautista:2020ahg} the covariance matrix is
\begin{equation}
   \textrm{Cov}_\textrm{eBOSS-LRG}=
   \begin{pmatrix}
       0.1076634 & -0.0583182 \\
       -0.0583182 & 0.2838176
\end{pmatrix}
\end{equation}

whereas for the eBOSS Quasar data \cite{Hou:2020rse,Neveux:2020voa} 
\begin{equation}
   \textrm{Cov}_\textrm{eBOSS-Quasar}=
  \begin{pmatrix}
    0.63731604 & 0.1706891 \\
     0.1706891 & 0.30468415
\end{pmatrix}.
\end{equation}

Finally the covariance matrix for the Ly$\alpha$ forest data is given by
\begin{equation}
   \textrm{Cov}_{\textrm{Ly}\alpha}=
   \begin{pmatrix}
      1.3225   & -0.1009     \\
      -0.1009  &  0.0380 
   \end{pmatrix} .
\end{equation}

\begin{table}
\caption{BAO measurements.}
\begin{ruledtabular}
\begin{tabular}{ccc}
 $z_\textrm{eff}$                     &  Measurement                                          &   Reference    \\[+0mm]
 \hline \\[-2mm]
 $0.122$    & $D_V\left(r_{d,\textrm{fid}}/r_d\right)$ [Mpc]                                              $= 539\pm 17$ [Mpc]  &  \cite{Carter:2018vce} \\[+1mm]
  \hline \\[-2mm]
 $0.38$     & $D_M/r_d$                   $= 10.2341 \pm 0.1691$   &  \cite{Gil-Marin:2020bct} \\[+1mm]
 $0.38$     & $D_H/r_d$   $= 24.9806\pm 0.7285$   &  \cite{Gil-Marin:2020bct}  \\[+1mm]
 $0.51$     & $D_M/r_d$                    $= 13.3660 \pm 0.2037$  &  \cite{Gil-Marin:2020bct}  \\[+1mm]
 $0.51$     & $D_H/r_d$  $= 22.3166   \pm 0.5717$   &  \cite{Gil-Marin:2020bct}  \\[+1mm]

 \hline \\[-2mm]
 $0.698$    & $D_M/ r_d$                                               $= 17.8582\pm0.3281$      & \cite{Gil-Marin:2020bct,Bautista:2020ahg} \\[+1mm]
 $0.698$    & $D_H / r_d$                                               $= 19.3258\pm0.5327$     & \cite{Gil-Marin:2020bct,Bautista:2020ahg} \\[+1mm]
 
 \hline \\[-2mm]
 $0.835$     & $D_M/r_d$                   $= 18.92\pm 0.51$       & \cite{DES:2021wwk}  \\[+1mm]
 \hline \\[-2mm]
 $1.48$    & $D_M/ r_d$                                               $= 30.6876\pm 0.7983$      & \cite{Hou:2020rse,Neveux:2020voa} \\[+1mm]
 $1.48$    & $D_H / r_d$                                               $= 13.2609\pm 0.5520$     & \cite{Hou:2020rse,Neveux:2020voa} \\[+1mm]

 \hline \\[-2mm]
 $2.334$    & $D_M / r_d$                                               $= 37.5^{+1.2}_{-1.1}$      & \cite{duMasdesBourboux:2020pck} \\[+1mm]
 $2.334$    & $D_H / r_d$                                               $= 8.99^{+0.20}_{-0.19}$     & \cite{duMasdesBourboux:2020pck} \\[+0mm]
\end{tabular}
\\[+1mm]
\begin{flushleft}
Note: For the data point at $z = 0.122$ the sound horizon size (at the drag epoch), for the fiducial values, computed using \eqref{eqn:sound_horizon} is  $r_{d,\textrm{fid}}=148.11~\textrm{Mpc}$ \cite{Carter:2018vce}.
\end{flushleft}
\end{ruledtabular}
\label{tab:bao}
\end{table}
\textbf{Cosmic chronometers:} We include in our analysis 32 measurements of $H(z_i)$, covering the redshift range $0.07 \leq z \leq 1.965$, obtained with the differential age technique. For those data points that are correlated with each other (see \cite{Moresco:2020fbm} for more details) the corresponding covariance matrix can be computed with the script provided here \footnote{ \url{https://gitlab.com/mmoresco/CCcovariance/-/blob/master/examples/CC_covariance.ipynb}}. The list of measurements of the Hubble parameter can be seen in Table \ref{tab:Hubble_data}. Those that appear without the corresponding error bars are the ones that are correlated with each other.  \\

\begin{table}
\caption{Hubble parameter data.}
\begin{ruledtabular}
\begin{tabular}{lccc}
  $z$     & $H(z)$                     &    Reference  \\[+0mm]
          & (km s$^{-1}$ Mpc$^{-1}$)   &               \\[+0mm]
  \hline \\[-2mm]
$0.07$    & $69.0 \pm 19.6$    & \cite{Zhang:2012mp}  \\ 
$0.09$    & $69.0 \pm 12.0$    & \cite{Simon:2004tf}  \\
$0.12$    & $68.6 \pm 26.2$    & \cite{Zhang:2012mp}  \\
$0.17$    & $83.0 \pm 8.0$     & \cite{Simon:2004tf}  \\
$0.2$     & $72.9 \pm 29.6$    & \cite{Zhang:2012mp}  \\
$0.27$    & $77.0 \pm 14.0$    & \cite{Simon:2004tf}  \\
$0.28$    & $88.8 \pm 36.6$    & \cite{Zhang:2012mp}  \\
$0.4$     & $95.0 \pm 17.0$    & \cite{Simon:2004tf}  \\
$0.47$    & $89.0 \pm 50.0$    & \cite{Ratsimbazafy:2017vga}  \\
$0.48$    & $97.0 \pm 62.0$    & \cite{Stern:2009ep}  \\
$0.75$    & $98.8 \pm 33.6$    & \cite{Borghi:2021rft}  \\
$0.88$    & $90.0 \pm 40.0$    & \cite{Stern:2009ep}  \\
$0.9$     & $117.0 \pm 23.0$   & \cite{Simon:2004tf}  \\
$1.3$     & $168.0 \pm 17.0$   & \cite{Simon:2004tf}  \\
$1.43$    & $177.0 \pm 18.0$   & \cite{Simon:2004tf}  \\
$1.53$    & $140.0 \pm 14.0$   & \cite{Simon:2004tf}  \\
$1.75$    & $202.0 \pm 40.0$   & \cite{Simon:2004tf}  \\
$0.1791$  & $74.91$            & \cite{Moresco:2020fbm}  \\
$0.1993$  & $74.96$            & \cite{Moresco:2020fbm}  \\
$0.3519$  & $82.78$            & \cite{Moresco:2020fbm}  \\
$0.3802$  & $83.0$             & \cite{Moresco:2020fbm}  \\
$0.4004$  & $76.97$            & \cite{Moresco:2020fbm}  \\
$0.4247$  & $87.08$            & \cite{Moresco:2020fbm}  \\
$0.4497$  & $92.78$            & \cite{Moresco:2020fbm}  \\
$0.4783$  & $80.91$            & \cite{Moresco:2020fbm}  \\
$0.5929$  & $103.8$            & \cite{Moresco:2020fbm}  \\
$0.6797$  & $91.6$             & \cite{Moresco:2020fbm}  \\
$0.7812$  & $104.5$            & \cite{Moresco:2020fbm}  \\
$0.8754$  & $125.1$            & \cite{Moresco:2020fbm}  \\
$1.037$   & $153.7$            & \cite{Moresco:2020fbm}  \\
$1.363$   & $160.0$            & \cite{Moresco:2020fbm}  \\
$1.965$   & $186.5$            & \cite{Moresco:2020fbm}
\end{tabular}
\end{ruledtabular}
\label{tab:Hubble_data}
\end{table}

\textbf{SNIa:} We employ the data from the Pantheon+\&SH0ES compilation \cite{Scolnic:2021amr,Brout:2022vxf} which contains 1701 light curves obtained from 1550 distinct SNIa with redshifts in the range $0.001\leq z\leq 2.26$. Within these 1550 SNIa there are 42 SNIa (with 77 light curves associated) in 37 Cepheid host galaxies, whose distance moduli have been obtained by measuring the characteristic period-luminosity relation of these variable stars. These measurements have been carried out by the the SH0ES collaboration \cite{Riess:2021jrx}. The fact of using the 77 distance moduli provided by the SH0ES team turns out to be fundamental to break the well-known degeneracy between the Hubble parameter $H_0$ and the absolute magnitude, which here we denote by $M$. Let us provide some specifics of the formulation used in the corresponding likelihood. The basic observable is the apparent magnitude $m$ which can be expressed as follows
\begin{equation}
m(z_{\textrm{HD}},z_{\textrm{HEL}}) = \mu(z_{\textrm{HD}},z_{\textrm{HEL}}) + M,    
\end{equation}
being $\mu$ the distance modulus 
\begin{equation}\label{eqn:distance_modulus}
\mu(z_{\textrm{HD}},z_{\textrm{HEL}}) = 25 + 5\textrm{log}_{10}\left(\frac{\tilde{d}_L(z_{\textrm{HD}},z_{\textrm{HEL}})}{1 \textrm{Mpc}}\right),    
\end{equation}
where the quantity $\tilde{d}_L(z_{\textrm{HD}},z_{\textrm{HEL}})$ can be understood as a corrected luminosity distance. Therefore we have: 
\begin{equation}
\tilde{d}_L(z_{\textrm{HD}},z_{\textrm{HEL}}) = \left(\frac{1+ z_{\textrm{HEL}}}{1+z_{\textrm{HD}}}\right)d_L(z_{\textrm{HD}}),    
\end{equation}
being the expression for the luminosity distance in a spatially-flat universe the usual one
\begin{equation}
d_L(z_{\textrm{HD}}) = \frac{(1+z_{\textrm{HD}})}{H_0}\int^{z_{\textrm{HD}}}_{0}\frac{dz^{\prime}}{E(z^{\prime})}.    
\end{equation}
In the above expressions $z_{\textrm{HD}}$ stands for the Hubble diagram redshift whereas $z_{\textrm{HEL}}$ represents the heliocentric redshift. \\
When the light curve considered is one of the 77 mentioned before (they are labeled with IS\_CALIBRATOR = 1 in the data file) instead of using \eqref{eqn:distance_modulus} to compute the distance modulus, we consider SH0ES' measurement for this quantity. This constraints greatly the value of the absolute magnitude $M$ (which in our analysis has the role of a nuisance parameter) thus breaking the existing degeneracy between $H_0-M$. \\

We consider the following joint $\chi^2$-function 
\begin{equation}\label{eq:chi2_total}
\chi^2_{\text{tot}} = \chi^2_{\text{CMB}} + \chi^2_{\text{BAO}} + \chi^2_{\text{H}} + \chi^2_{\text{SNIa}}   
\end{equation}
in order to study the performance of the $\Lambda$CDM and TDiff cosmological models when they are confronted with the data set previously described. The data set considered in \eqref{eq:chi2_total} will be referred to as Baseline. We explore the parameter-space that characterizes each of the models with the Markov chain Monte Carlo analysis, making use of the Metropolois-Hastings algorithm \cite{Metropolis:1953am,Hastings:1970aa}. Once the converged chains have been obtained we utilize the \texttt{GetDist} code \cite{getdist} to obtain the mean values of the different cosmological parameters, the associated confidence intervals and the posterior distributions.  When exploring the parameter-space we have set the following priors for the three common parameters $0.01 < \Omega_b < 0.1$, $40 < H_0[\text{km/s/Mpc}] < 100$ and $-20 < M < -18$. In the $\Lambda$CDM case we have considered $0.1 < \Omega_m < 0.5$ whereas for the TDiff model $1.7 < \beta < 2.3$. We want to compare the performance of the $\Lambda$CDM model and the TDiff model, therefore in this work we use the deviance information criterion (DIC) \cite{DIC}, whose value can be computed from 
\begin{equation}\label{eqn:DIC}
\textrm{DIC} = \chi^2(\bar{\theta}) + 2p_D\,.
\end{equation}
In the above expression,  $p_D = \overline{\chi^2} - \chi^2(\bar{\theta})$ stands for the effective number of parameters, $\overline{\chi^2}$ is the mean value of the $\chi^2$-function and $\bar{\theta}$ contains the mean value of the free parameters. We define the difference with respect to the $\Lambda$CDM model as follows
\begin{equation}\label{eq:DIC_difference}
\Delta\text{DIC} = \text{DIC}_{\Lambda\text{CDM}} - \text{DIC}_\text{TDiff},    
\end{equation}
as a consequence if we obtain positive values of $\Delta\text{DIC}$ means that the TDiff model is favored over the $\Lambda$CDM models whereas negative values indicate the other way around. Having defined the difference in the DIC values as in \eqref{eq:DIC_difference}, according to the usual standards if we find values $0 \leq \Delta\textrm{DIC}<2$ we say there is {\it weak} evidence in favor of the TDiff model. If we get values $2 \leq \Delta\textrm{DIC}<6$ we speak of {\it positive} evidence whereas values $6 \leq \Delta\textrm{DIC}<10$ point out to {\it strong} evidence in favor of the TDiff. Finally, finding values $\Delta\text{DIC}>10$ would allow us to claim {\it very strong} evidence. As we mentioned before negative values would indicate evidence in favor of the standard model. 

In order to complement the DIC we also make use of the Bayesian Evidence which has been computed following the procedure detailed in \cite{Heavens:2017hkr}. Here we define the Bayes' ratio as 
\begin{equation}
\mathcal{B} \equiv \frac{E_{\text{TDiff}}}{E_{\Lambda\text{{CDM}}}}
\end{equation}
where $E$ stands for the Bayesian Evidence. With the above definition, and as with the DIC, positive values indicate that the TDiff model is favored over the $\Lambda$CDM model with different levels of significance. Therefore if $\ln\mathcal{B}\leq 1$ there is {\it inconclusive} evidence, for $1<\ln\mathcal{B}\leq 2.5$ we get {\it moderate} evidence whereas if we obtain  $2.5<\ln\mathcal{B}\leq 5$ we speak of {\it strong} evidence. Finally if we find values $\ln\mathcal{B}>5$ we are allowed to claim {\it very strong} evidence in favor of the TDiff model. For negative values of $\ln\mathcal{B}$ is the other way around. 

\section{Results}\label{sect:Results}
\begin{figure*}[t!]
    \centering
    \includegraphics[scale=0.77]{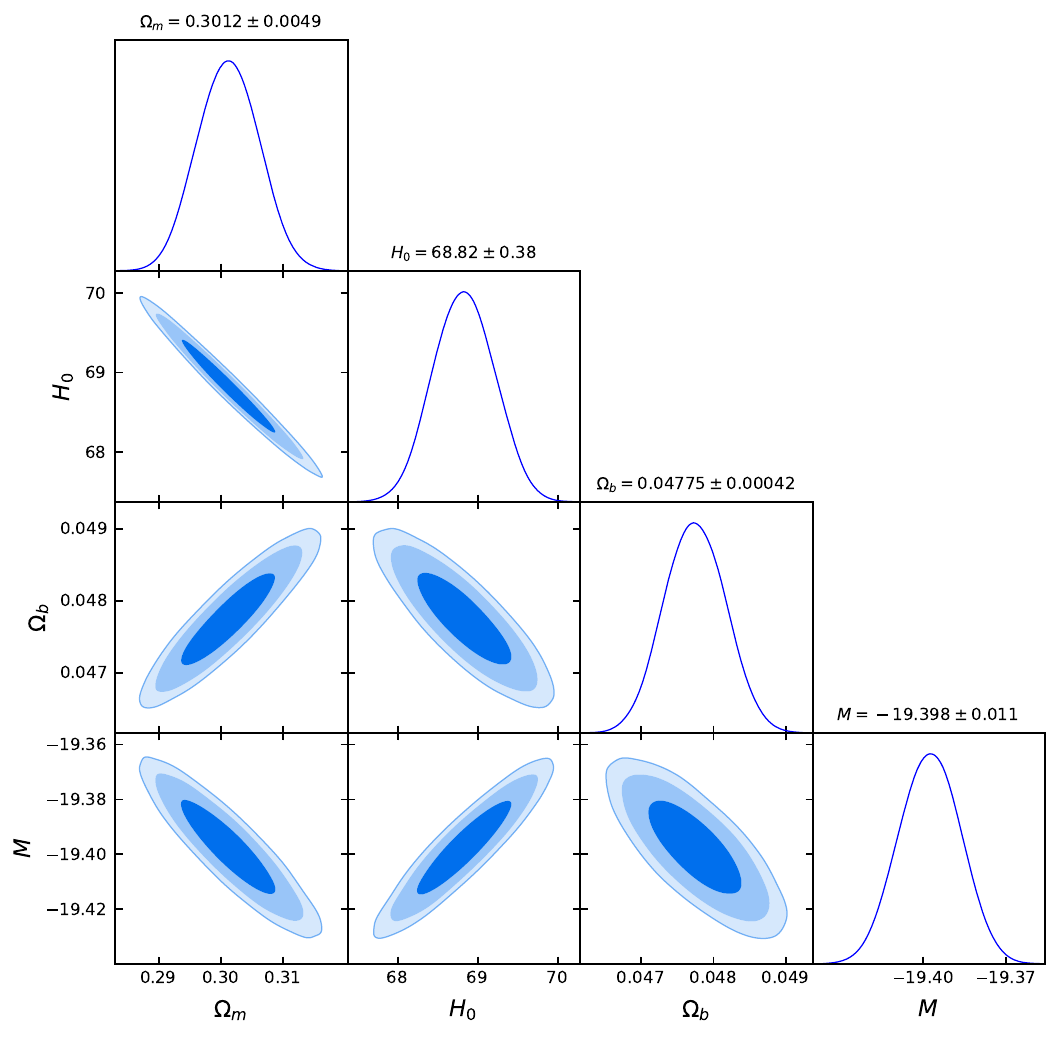}
    \caption{For the $\Lambda$CDM model contour plots at 1$\sigma$, 2$\sigma$ and 3$\sigma$ confidence level, as well as, the one-dimensional posterior distributions for the Baseline data set. The $H_0$ parameter is expressed in km/s/Mpc units. }
    \label{fig:LCDM_contour_plots}
\end{figure*}

\begin{figure*}[t!]
    \centering
    \includegraphics[scale=0.77]{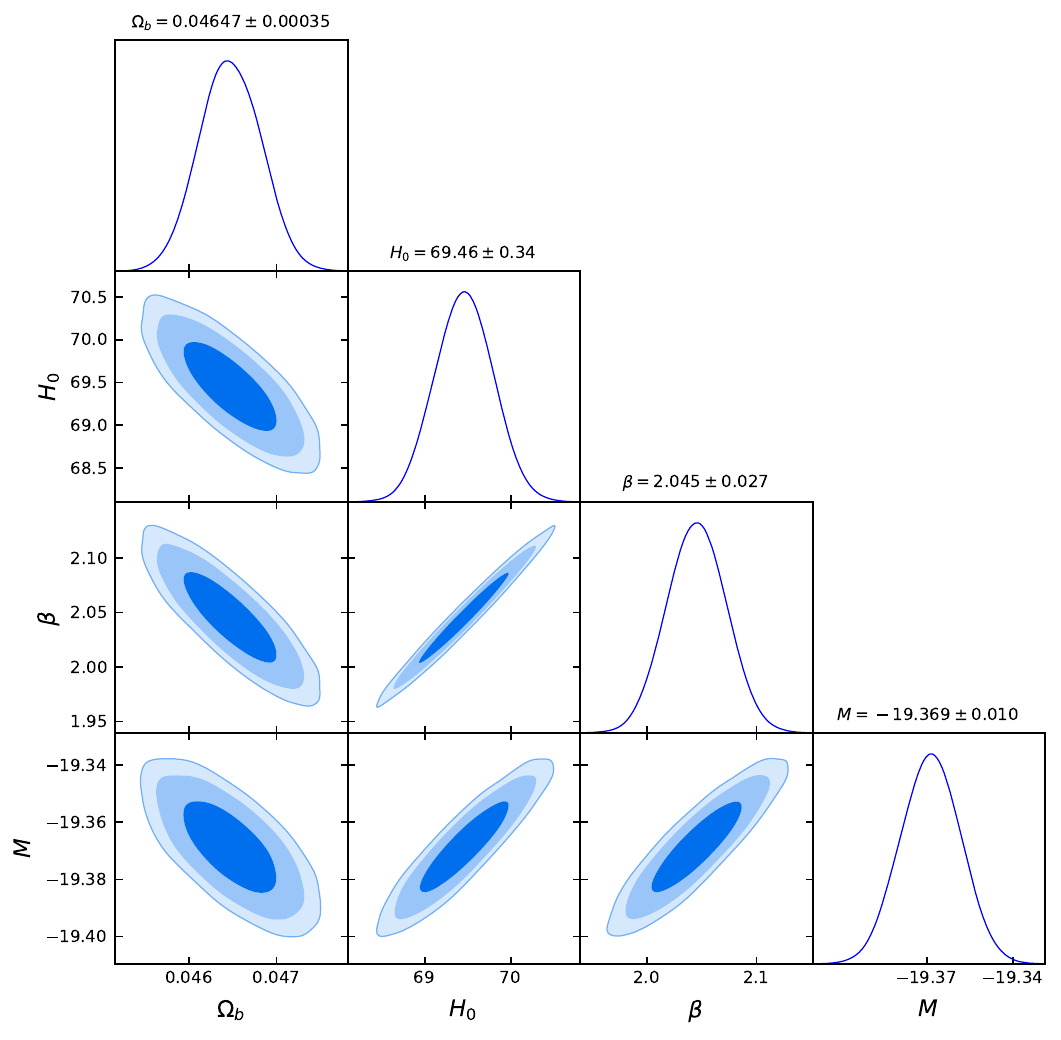}
    \caption{Same as in Fig. \ref{fig:LCDM_contour_plots} but for the TDiff model. }
    \label{fig:TDiff_contour_plots}
\end{figure*}

The best-fit values and the corresponding 68 \% confidence intervals, obtained after considering the Baseline data set (see Eq. \eqref{eq:chi2_total}), can be found in Table \ref{tab:main_results}  and in Figs. \ref{fig:LCDM_contour_plots} and \ref{fig:TDiff_contour_plots}.
We see that the TDiff model provides an excellent fit to the considered data set with a minimum $\chi^2$ below that of $\Lambda$CDM with 
same number of free parameters. As a matter of fact $\Delta\text{DIC}$ = 10.39 indicates {\it very strong} evidence whereas $\ln\mathcal{B}=4.02$ indicates {\it strong} evidence in favor of the TDiff model when it is confronted against the $\Lambda$CDM. The main difference appears in the mean value
of the Hubble parameter which is larger than in the $\Lambda$CDM
and with similar error bars. As we will show below, this will mainly explain the better fit to the complete data set.  

As commented before, at high redshift, the TDiff scalar field mimics the non-relativistic matter behaviour as it is clear from \eqref{eq:Omega_phi_a_small} and consequently the effective matter density parameter at that time  
is $\Omega^{\text{eff}}_M =\Omega^{\text{eff}}_{\text{DM}} + \Omega_B = 0.3005\pm 0.0043$ (computed as a derived parameter obtained with the Baseline data set), which can be compared with the $\Lambda$CDM value $\Omega_M = 0.3012\pm 0.0049$. Therefore, the behaviour of the TDiff model is very similar to the $\Lambda$CDM model at high-redshift, when the contribution from dark energy is negligible. 

On the other hand if we want to study the behaviour of the TDiff model at late times, we may look at the equation of state parameter, which at present time, takes the value $\omega_\phi(z=0) = -0.6746\pm 0.0029$. The equivalent quantity for the $\Lambda$CDM model would be $\omega_{\Lambda\text{CDM}}(z=0)  = -0.7383\pm 0.0044$ with $\omega_{\Lambda\text{CDM}}$ defined in \eqref{wLCDM}. The low-redshift evolution of both parameters can be seen in Fig. \ref{fig:eqn_state}. The fact that the value in the $\Lambda$CDM model is, in absolute value, greater than the value in the TDiff model indicates that the behaviour of the scalar field is quintessence-like. This is confirmed thanks to the effective equation of state parameter for the dark energy component defined in Eq. \eqref{eq:w_DE_TDiff}, whose present value $\omega^{\text{eff}}_{\textrm{DE}}(z=0)=-0.9156\pm 0.0016$ lies in the quintessence region. As shown in Fig. \ref{fig:w_DE_TDiff}, the TDiff model evolves from a phantom effective dark energy equation of state $\omega^{\text{eff}}_{\textrm{DE}}(z)=-2$ at high redshift to a 
quintessence-like value at present.  

We also find some differences between the models when it comes to the redshift value that marks the transition from a decelerated to an  accelerated expanding universe. Whereas for the $\Lambda$CDM we have $z_t\simeq 0.668$ for the TDiff we get $z_t\simeq 0.794$ and the corresponding value of the $b(a)$ parameter $b_t\simeq 3.1314$.    
In Fig. \ref{fig:b_param} we show the behaviour of $b(z)$ which is constant in the matter dominated era, being this constant value $b(z\gg 1)\simeq \sqrt{1+2\beta}e^{\frac{\beta}{2}}\simeq 6.272$.

\subsection{The $H_0$-tension}
The persisting mismatch between the value obtained by the Planck collaboration ($H_0=67.36\pm 0.54$ km/s/Mpc from TT,TE,EE+lowE+lensing data \cite{Planck:2018vyg}) and the SH0ES measurement ($H_0=73.04\pm 1.04$ km/s/Mpc \cite{Riess:2021jrx} which hereafter will be denoted by $H^{\textrm{SH0ES}}_0$), obtained with the inverse cosmic distance ladder method, has become a central topic of study in cosmology in recent years. In order to calculate the tension between two given values we utilize the following estimator
\begin{equation}
T_{H_0} = \frac{|H^{(1)}_0 - H^{(2)}_0|}{\sqrt{\sigma^2_{(1)} + \sigma^2_{(2)} }}.     
\end{equation}
Therefore the tension between the mentioned values reaches the astonishing level of $T_{H_0}\simeq 4.85\sigma$. At this stage it is still not clear whether the tension is somehow caused by some sort of unknown systematic or if it is actually the clearest hint we have so far of physics beyond the standard model. 

This section is specifically aimed to check how the TDiff model deals with the $H_0$-tension. It has been argued \cite{Keeley:2022ojz} that no low-redshift solution can completely solve the tension. The authors consider a very flexible parameterization of the equation of the state parameter to prove that the 1-dimensional posterior distribution of the $H_0$ parameter ($H_0=68.08\pm 0.97$ km/s/Mpc), obtained from a CMB+BAO+SNIa (the SNIa data do not contain SH0ES measurement unlike the one we consider here) data set, does not overlap with SH0ES' measurement posterior distribution. 

For the sake of comparison we have tested the TDiff model with a combination of data CMB+BAO+SNIa$^{*}$, where the $*$ indicates that in this occasion the SNIa data do not contain the SH0ES contribution. Regarding the Hubble parameter we get the following result $H_0=68.95\pm 0.37$ km/s/Mpc, which is perfectly compatible with the value provided in \cite{Keeley:2022ojz}.

\begin{figure*}[t!]
    \centering
    \includegraphics[scale=0.7]{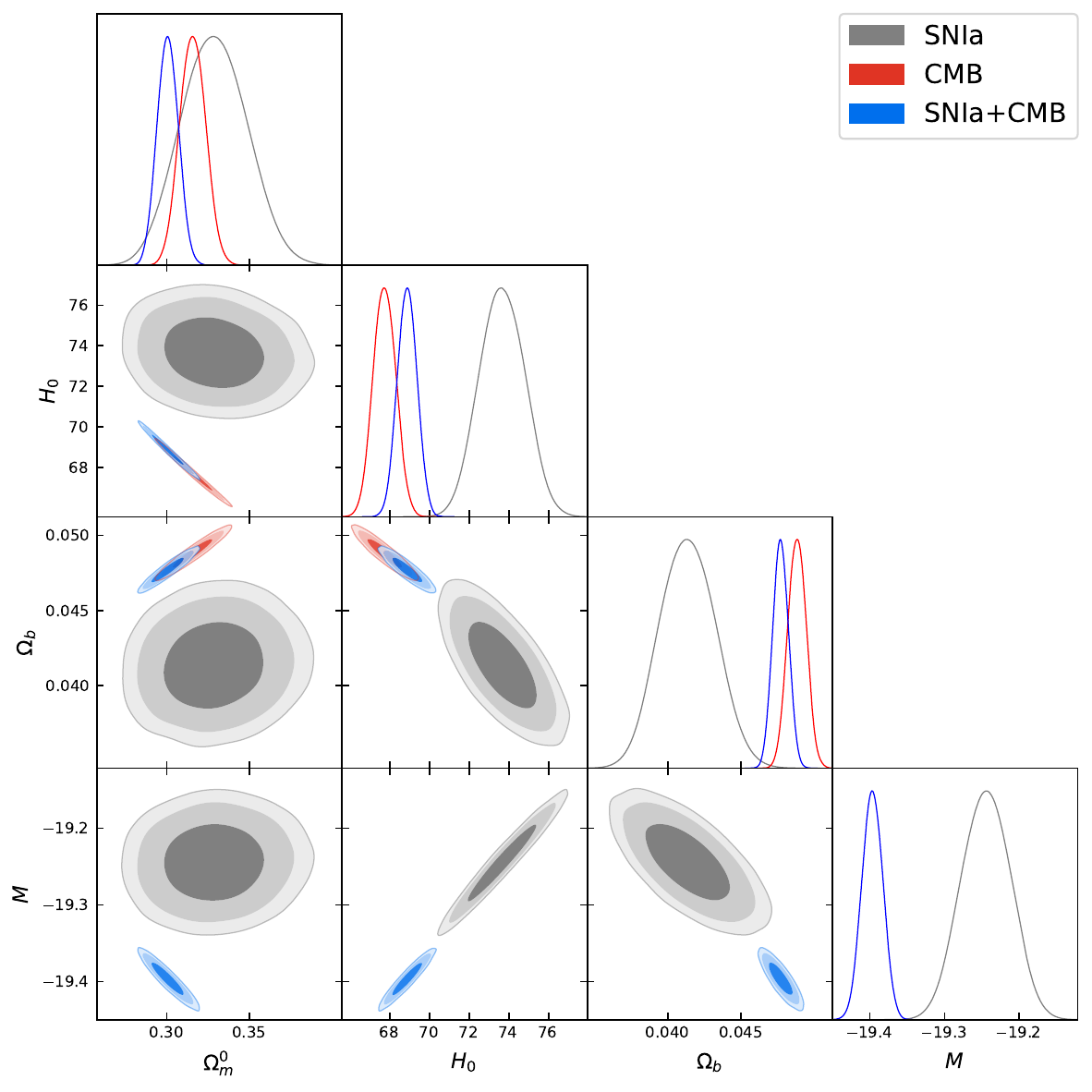}
    \caption{For the $\Lambda$CDM model contour plots at 1$\sigma$, 2$\sigma$ and 3$\sigma$ confidence level for the SNIa + prior on $\omega_b$, CMB and SNIa+CMB data. The $H_0$ parameter is expressed in km/s/Mpc units. }
    \label{fig:LCDM_CMB_SNIa}
\end{figure*}

\begin{figure*}[t!]
    \centering
    \includegraphics[scale=0.7]{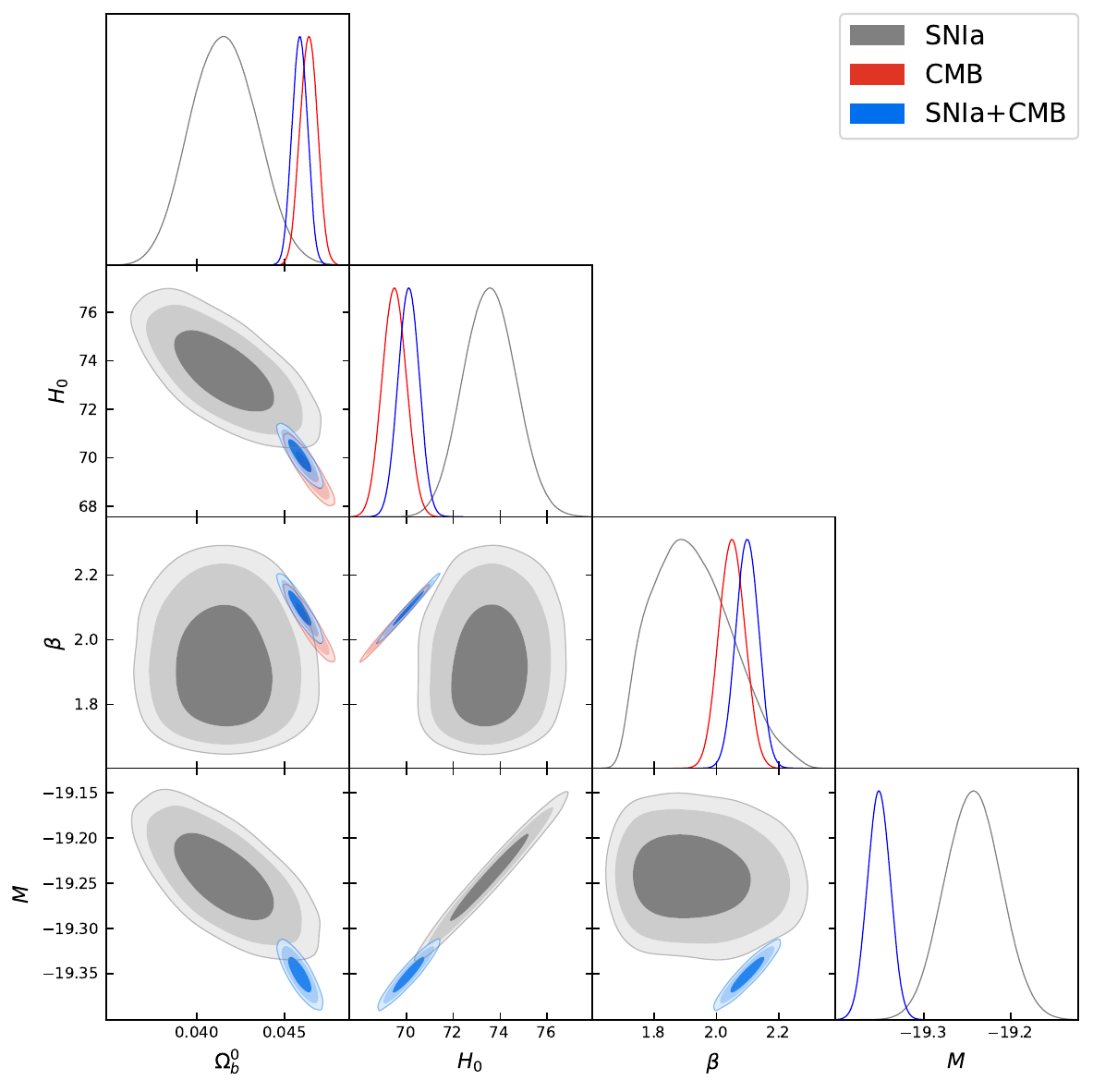}
    \caption{Same as in Fig.\ref{fig:LCDM_CMB_SNIa} but for the TDiff model. }
    \label{fig:TDiff_CMB_SNIa}
\end{figure*}

If we have a look at Tables \ref{tab:CMB_results} and \ref{tab:SNIa_results} we can see the results obtained when first only the CMB and then only the SNIa data (in addition we consider a model-independent BBN prior for the density of baryons $\omega_b = 0.02244\pm 0.00069$ \cite{ParticleDataGroup:2022pth} determined by D/H abundances) are considered, whereas in Table \ref{tab:CMB_and_SNIa_results} we provide the results when both of these data sets are jointly considered. In Table \ref{tab:CMB_results} we can see that for the TDiff model we get $H_0=69.49\pm 0.51$ km/s/Mpc which is in 3.1$\sigma$ tension with $H^{\textrm{SH0ES}}_0$, representing a clear reduction with respect to the result obtained for the $\Lambda$CDM ($H_0 = 67.72\pm 0.57$ km/s/Mpc showing a discrepancy of 4.5$\sigma$ with the SH0ES measurement). On the other hand, when only the likelihood for the SNIa data is taken into account we get almost the same results for both models (see Table \ref{tab:SNIa_results}). It is when we analyze both data sets together (results provided in Table \ref{tab:CMB_and_SNIa_results}) when we are able to appreciate the different performance of the two models when it comes to fitting the data. We obtain $\Delta$DIC=15.80 and $\ln\mathcal{B}=6.90$ which according to the statistical standards indicates a {\it very strong} evidence in favor of the TDiff model when it is compared with the standard model. In Figs. \ref{fig:LCDM_CMB_SNIa} and \ref{fig:TDiff_CMB_SNIa} we display the contour plots for the $\Lambda$CDM and the TDiff models respectively. While for the standard model there are panels where the contours do not overlap at more than 3$\sigma$ c.l., which is of course a reflection of the previously presented $H_0$-tension, in the case of the TDiff we observe that the model shows greater capacity to fit both data sets, CMB and SNIa, simultaneously. It is worth noticing that the TDiff model can alleviate the $H_0$-tension without including extra parameters, with respect to the $\Lambda$CDM model, which normally increases the size of the error bars thus reducing the tension. In fact, they are smaller than in the standard model which means that the softening of the tension actually comes from an increase of the mean value. Additionally, this goes hand-in-hand with a fit of observational data comparable, if not better, to the one of the $\Lambda$CDM model, something that despite the number of cosmological models studied in the literature \cite{DiValentino:2021izs}, is not easy to achieve. 

Even in the light of these promising results we are not claiming that the $H_0$-tension is solved within the context of the TDiff cosmological model. When CMB data are considered the level of tension, with respect to the local measurement $H^{\textrm{SH0ES}}_0$, is still $\sim 3\sigma$ which although it is not as alarming as the $\sim 5\sigma$ tension found for the $\Lambda$CDM when the full Planck 2018 TT,TE,EE+lowE+lensing likelihood is analyzed, it is still quite high and consequently it cannot be ignored. It is also important to remember that in this work we have used an approximation for the CMB likelihood (see \eqref{eq:CMB_data_points} and \eqref{eq:correlationCMB}) therefore some variations in the cosmological parameter constraints may be expected when considering the full likelihood, something that the authors are currently studying. All in all, we find that the TDiff model presents an interesting flexibility when it comes to push up the value of the $H_0$ parameter, thus loosening the discrepancy with the $H^{\textrm{SH0ES}}_0$ measurement but more detailed studies are needed in order to confirm these results. 


\begin{table}[h!]
\centering
\begin{tabular}{ccc}
\multicolumn{3}{c}{Baseline}
\\\hline
                               & $\Lambda$CDM           & TDiff            \\ \hline
\multicolumn{1}{c}{$\Omega_B$}     & $ 0.04775\pm 0.00042$  &  $0.04647\pm 0.00035$               \\ \hline
\multicolumn{1}{c}{$\Omega_M$} & $0.3012\pm 0.0049$              & -  \\ \hline
\multicolumn{1}{c}{$H_0$ [km/s/Mpc]} & $68.82\pm 0.38$ & $69.46\pm 0.34$ \\ \hline
\multicolumn{1}{c}{$\beta$}        & -  & $2.045\pm 0.027$   \\ \hline  
\multicolumn{1}{c}{$M$}        & $-19.398\pm 0.011$  & $-19.369\pm 0.010$   \\ \hline  \\[-9pt]
\multicolumn{1}{c}{$\small{\chi^2_{\text{min}}}$}   & $1338.34$   & $1327.97$ \\ \hline
\multicolumn{1}{c}{$\small{\Delta \text{DIC}}$}   & -   & $10.39$ \\ \hline
\multicolumn{1}{c}{$\small{\ln\mathcal{B}}$}   & -   & $4.02$ \\ \hline\hline
\end{tabular}
\caption{Mean values with the 68\% confidence interval for the cosmological parameters considered obtained by analyzing the Baseline data set (see Eq. \eqref{eq:chi2_total}). We also show the minimum value of the $\chi^2$-function, the incremental value of the DIC and the value of $\ln\mathcal{B}$.}
\label{tab:main_results}
\end{table}

\begin{table}[h!]
\centering
\begin{tabular}{ccc}
\multicolumn{3}{c}{CMB}
\\\hline
                               & $\Lambda$CDM           & TDiff            \\ \hline
\multicolumn{1}{c}{$\Omega_B$}     & $ 0.04882\pm 0.00062$  &  $0.04638\pm 0.00049$               \\ \hline
\multicolumn{1}{c}{$\Omega_M$} & $0.3158\pm 0.0078$              & -  \\ \hline
\multicolumn{1}{c}{$H_0$ [km/s/Mpc]} & $67.72\pm 0.57$ & $69.49\pm 0.51$ \\ \hline
\multicolumn{1}{c}{$\beta$}        & -  & $2.049\pm 0.041$   \\ \hline  \\[-9pt]
\multicolumn{1}{c}{$\small{\chi^2_{\text{min}}}$}   & $0$   & $0$ \\ \hline
\multicolumn{1}{c}{$\small{\Delta \text{DIC}}$}   & -   & $-0.97$ \\ \hline
\multicolumn{1}{c}{$\small{\ln\mathcal{B}}$}   & -   & $-0.09$ \\ \hline\hline
\end{tabular}
\caption{Mean values with the 68\% confidence interval for the cosmological parameters when only CMB data are considered. }
\label{tab:CMB_results}
\end{table}


\begin{table}[h!]
\centering
\begin{tabular}{ccc}
\multicolumn{3}{c}{SNIa+prior on $\omega_b$}
\\\hline
                               & $\Lambda$CDM           & TDiff            \\ \hline
\multicolumn{1}{c}{$\Omega_B$}     & $ 0.0414\pm 0.0019$  &  $0.0415\pm 0.0019$               \\ \hline
\multicolumn{1}{c}{$\Omega_M$} & $0.329\pm 0.020$              & -  \\ \hline
\multicolumn{1}{c}{$H_0$ [km/s/Mpc]} & $73.7\pm 1.1$ & $73.6\pm 1.1$ \\ \hline
\multicolumn{1}{c}{$\beta$}        & -  & $1.92^{+0.10}_{-0.15}$   \\ \hline
\multicolumn{1}{c}{$M$}     & $ -19.244\pm 0.033$  &  $-19.243\pm 0.032$               \\ \hline\\[-9pt]
\multicolumn{1}{c}{$\small{\chi^2_{\text{min}}}$}   & $1277.13$   & $1277.48$ \\ \hline
\multicolumn{1}{c}{$\small{\Delta \text{DIC}}$}   & -   & $0.23$ \\ \hline
\multicolumn{1}{c}{$\small{\ln\mathcal{B}}$}   & -   & $-0.16$ \\ \hline\hline
\end{tabular}
\caption{Mean values with the 68\% confidence interval for the cosmological parameters when only SNIa data and the prior $\omega_b = 0.02244\pm 0.00069$ \cite{ParticleDataGroup:2022pth} are considered. }
\label{tab:SNIa_results}
\end{table}  


\begin{table}[h!]
\centering
\begin{tabular}{ccc}
\multicolumn{3}{c}{CMB+SNIa}
\\\hline
                               & $\Lambda$CDM           & TDiff            \\ \hline
\multicolumn{1}{c}{$\Omega_B$}     & $ 0.04770\pm 0.00053$  &  $0.04587\pm 0.00044$               \\ \hline
\multicolumn{1}{c}{$\Omega_M$} & $0.3007\pm 0.0064$              & -  \\ \hline
\multicolumn{1}{c}{$H_0$ [km/s/Mpc]} & $68.86\pm 0.49$ & $70.10\pm 0.45$ \\ \hline
\multicolumn{1}{c}{$\beta$}        & -  & $2.098\pm 0.036$   \\ \hline
\multicolumn{1}{c}{$M$}        & $-19.396\pm 0.014$  & $-19.351\pm 0.013$   \\ \hline  \\[-9pt]
\multicolumn{1}{c}{$\small{\chi^2_{\text{min}}}$}   & $1310.06$   & $1294.13$ \\ \hline
\multicolumn{1}{c}{$\small{\Delta \text{DIC}}$}   & -   & $15.80$ \\ \hline
\multicolumn{1}{c}{$\small{\ln\mathcal{B}}$}   & -   & $6.90$ \\ \hline\hline
\end{tabular}
\caption{Mean values with the 68\% confidence interval for the cosmological parameters when SNIa+CMB data are considered. }
\label{tab:CMB_and_SNIa_results}
\end{table} 


\section{Conclusions}\label{sect:Conclusions}

In this work we have presented a unified model for the dark sector with a simple canonical kinetic term in the action. This theory is based on a scalar field $\phi$ that explicitly breaks the full diffeomorphism invariance of the action down to transverse diffeomorphisms i.e. the action \eqref{eq:action_with_kinetic_term} is invariant under coordinate transformations that do not change the determinant of the metric tensor. The function characterizing the Diff breaking is 
a simple exponential $f(g)=e^{-\beta g}$. The model has the same number of free parameters as $\Lambda$CDM, being the common ones $\Omega_B$, $H_0$ and the nuisance parameter $M$ and where the $\beta$ parameter replaces  the $\Omega_M$ parameter of $\Lambda$CDM. 

Unlike  General Relativity, in the TDiff case there are two physical scale factors in the metric tensor at the background level, namely: $a(\tau)$ and $b(\tau)$, being $\tau$ a privileged time, established by the TDiff symmetry and related with the cosmological time $t$ through $b = dt/d\tau$. This is so because no time reparametrizations with fixed spatial coordinates (to maintain homogeneity and isotropy) are allowed by the symmetry. Even at the background level, the TDiff time $\tau$ differs from the time $t$ that a comoving observer would measure, so that for a process that takes a fixed $\Delta \tau$ amount of time, since $b$ decreases with cosmological time $t$, the corresponding $\Delta t$ becomes smaller. This can be thought of as some kind of time dilation of $\tau$ with respect to $t$.  

As stated previously we consider a unified fluid playing the role of both dark matter and dark energy, therefore, the expression chosen for $f(g)$ is such that the equation of state parameter \eqref{eqn:wphi} of the fluid interpolates between a matter-type behaviour and a cosmological constant-type one. In fact, at the background level we can effectively separate the contribution from the scalar field into two components, one mimicking the evolution of the non-relativistic matter and another one behaving as if it were a dark energy component which evolves from a phantom phase in the early universe to a quintessence-like behaviour at present. Other forms of the $f(g)$ function can be analyzed, but the simple exponential expression considered in this work already shows the flexibility of TDiff models to describe the dark sector without introducing additional parameters. 

Previous works \cite{transvsobs, Maroto:2023toq,Jaramillo-Garrido:2023cor,Bello-Morales:2023btf} aim at presenting viable models on cosmological backgrounds with this reduced TDiff symmetry, however, this is the first time that a consistent model is actually compared with experiment and constraints on its parameters are presented. We have considered different data sets in order to test the model in different scenarios. The results obtained after analyzing the Baseline data set, composed by the chain of data CMB+BAO+$H(z)$+SNIa, can be seen in Table \ref{tab:main_results} and in Figs. \ref{fig:LCDM_contour_plots} and \ref{fig:TDiff_contour_plots}. Encouraging signals are gathered indicating that the TDiff model can surpass the performance of the standard $\Lambda$CDM model  when it comes to fitting the data. However, these promising results are to be confirmed by including in the analysis the full likelihood for the CMB data and by considering the perturbation equations. 
The potential modification in the evolution of perturbations could affect some of the observables considered in the analysis. In particular, the CMB distance priors and BAO scales could be affected a priori. Those observables are 
determined by the acoustic scale at the time of recombination (drag epoch), however at high redshift, the TDiff scalar behaves exactly as a matter fluid, so that we do not expect any modification in the evolution of perturbations in the radiation or matter eras with respect to $\Lambda$CDM. Nevertheless, a confrontation with growth rate data or the full matter power spectrum would require a more detailed determination of the evolution of perturbations at low redshift which could deviate from $\Lambda$CDM.

We have also delved into the study of the well-known $H_0$-tension. Interestingly we observe that the TDiff model is able to alleviate the tension, when compared with the $\Lambda$CDM model, not stretching the error bars but rather pushing the mean value of Hubble parameter towards higher values. 

In light of the results obtained in this work we conclude that the general class of transverse diffeomorphism invariant models not only present an appealing theoretical framework, where there may be a way out of the cosmological constant problem, but also provides an interesting alternative to explain the cosmological data and what is more to alleviate some tensions that affect the $\Lambda$CDM model.

\acknowledgements{JdCP is supported by the Margarita
Salas fellowship funded by the European Union (NextGenerationEU). This work has been supported by the MINECO (Spain) project PID2019-107394GB-I00 (AEI/FEDER, UE).}

\section{Appendix: TDiff invariant scalars}

Let us  consider  an infinitesimal general coordinate transformation 
\begin{equation}
    \hat{x}^{\mu}=x^{\mu}+\xi^{\mu}(x),
\end{equation}
The corresponding transformation law for the metric tensor reads
\begin{equation}
    \hat{g}_{\mu\nu}(\hat{x})=\frac{\partial x^{\alpha}}{\partial \hat{x}^{\mu}}\frac{\partial x^{\beta}}{\partial \hat{x}^{\nu}}g_{\alpha\beta}(x),
\end{equation}
so that its determinant transforms as a tensor density
\begin{equation}
    \hat{g}(\hat{x})={\Bigg |}\text{det}\left(\frac{\partial x^{\alpha}}{\partial \hat{x}^{\mu}}\right){\Bigg |}^2g(x). 
\end{equation}
At an infinitesimal level, by using the identity $\text{det}(\mathbb{1}+A)=1+\text{tr}(A)+\mathcal{O}(A^2)$, we get
{\small\begin{equation}
    {\Bigg |}\text{det}\left(\frac{\partial x^{\alpha}}{\partial \hat{x}^{\mu}}\right){\Bigg |}^2= {\Bigg|}\text{det}\left(\delta^{\alpha}_{\mu}-\frac{\partial \xi^{\alpha}}{\partial \hat{x}^{\mu}}\right){\Bigg|}^2=1-2\partial_{\mu}\xi^{\mu}(x)+\mathcal{O}(\xi^2),
\end{equation}}where we have also made use of the fact that the derivatives of $\xi^{\mu}(x)$ with respect to the new coordinates $\hat{x}^{\mu}$ can be replaced with the derivatives with respect to the old coordinates to linear order in $\xi^{\mu}(x)$. This means that since
\begin{align}
    \hat{g}(\hat{x})=(1-2\partial_{\mu}\xi^{\mu})\,g(x), \\
    d^4\hat{x}=(1+\partial_{\mu}\xi^{\mu})\,d^4x,
\end{align}
for any coordinate transformation that satisfies
\begin{align}
    \partial_{\mu}\xi^{\mu}=0,
    \label{eqn:TDiff}
\end{align}
then any action term of the form
\begin{equation}
    S = \int d^4x\, f(g)\,\lagr,
\end{equation}
where $f(g)$ is an arbitrary function of the metric determinant and $\lagr$ is a scalar function of the matter fields, its derivatives and the metric tensor, is invariant under these transformations. Coordinate transformations that satisfy the condition \eqref{eqn:TDiff} are referred to as transverse diffeomorphisms.

Thus, to lowest order in metric derivatives,  the most general TDiff invariant action up to quadratic terms in derivatives of a real scalar field reads \cite{transvsobs,Maroto:2023toq}
{\small\begin{equation}
    S_{\phi} = \int d^4x\, \lagr = \int d^4x \left(\frac{1}{2}f_K(g)g^{\mu\nu}\partial_{\mu}\phi\partial_{\nu}\phi-f_V(g)V(\phi)\right),
    \label{eqn:Sgeneral}
\end{equation}}where $f_K(g)$ and $f_V(g)$ are arbitrary functions of the metric determinant. Notice that in principle, other (non-minimal) terms could be included involving derivatives of the metric tensor, but we will limit ourselves to the simplest minimal couplings.

\bibliography{referencias}
\end{document}